\documentclass[aps,prd,groupedaddress]{revtex4}
\usepackage{amsmath}
\usepackage{amssymb}
\usepackage{bbm}
\usepackage{epsfig}
\begin{document}
% versão referee
\title{Lepton flavour violating processes at the International Linear Collider}

\author{P.M. Ferreira$^{1}$~\footnote{ferreira@cii.fc.ul.pt}, R.B.
Guedes$^{1}$~\footnote{renato@cii.fc.ul.pt} and R.
Santos$^{1,2}$~\footnote{rsantos@cii.fc.ul.pt}}
\affiliation{$^{1}$ Centro de F\'{\i}sica Te\'orica e
Computacional, Faculdade de Ci\^encias, Universidade de Lisboa,
Avenida Professor Gama Pinto, 2, 1649-003
Lisboa, Portugal \\
$^{2}$ Department of Physics, Royal Holloway, University of
London, Egham, Surrey TW20 0EX United Kingdom }

\date{November, 2006}

\begin{abstract}
We study the effects of dimension six effective operators on the
flavour violating production and decay of leptons at the
International Linear Collider. Analytic expressions for the cross
sections, decay widths and asymmetries of all flavour changing
processes will be presented, as well as an analysis of the
feasibility of their observation at the ILC.
\end{abstract}

\maketitle

\section{Introduction}

In the next years the Large Hadron Collider (LHC) will start to
function and provide the scientific community with a new tool with
which to explore hitherto unknown regions of particle physics. We
expect many exciting discoveries to arise from LHC experiments.
However, the LHC is a hadronic machine, and as such precision
measurements will be quite hard to undertake there. Also, the
existence of immense backgrounds at the LHC may hinder discoveries
of new physical phenomena already possible at the energies that
this accelerator will achieve. Thus it has been proposed to build
a new electron-positron collider, the International Linear
Collider~\cite{ILC}. This would be a collider with energies on the
TeV range, with extremely high luminosities. The potential for new
physics with such a machine is immense. In this paper, we will
focus on a specific sector: the possibility of processes which
violate lepton flavour occurring.

We now know that the solar and atmospheric neutrino
problems~\cite{neut} arise, not from shortcomings of solar models,
but from particle physics. Namely, the recent findings by the SNO
collaboration~\cite{snow} have shown beyond doubt that neutrinos
oscillate between families as they propagate over long distances.
Leptonic flavour violation (LFV) is therefore an established
experimental fact. The simplest explanation for neutrino
oscillations is that neutrinos have masses different from zero -
extremely low masses, but non-zero nonetheless. Oscillations with
zero neutrino masses are possible, but only in esoteric
models~\cite{barr}. With non-zero neutrino masses, flavour violation
in the charged leptonic sector becomes a reality (whereas with
massless neutrinos, it is not allowed in the Standard Model (SM)).
This is a sector of particle physics for which we already have many
experimental results~\cite{pdg}, which set stringent limits on the
extent of flavour violation that may occur. Nevertheless, as we will
show in this paper, even with all known experimental constraints it
is possible that signals of LFV may be observed at the ILC, taking
advantage of the large luminosities planned for that machine. There
has been much attention devoted to this subject. For instance, in
refs.~\cite{zes} effective operators were used to describe LFV
decays of the $Z$ boson. LFV decays of the $Z$ boson were also
studied in many extensions of the SM~\cite{zesext}. The authors of
refs.~\cite{cann,ggilc} performed a detailed study of LFV at future
linear colliders, originating from Supersymmetric models. Finally, a
detailed study of the four fermion operators in the framework of LFV
is performed in~\cite{javier}. In that work the exact number of
independent four fermion operators is determined. Gauge invariance
is then used to constrain LFV processes which are poorly measured,
or not measured at all.

In this work we carry out a model-independent analysis of all
possible LFV interactions which might arise in extensions of the SM.
To do so, we utilise the effective operator formalism of
Buchm\"uller and Wyler~\cite{buch}, a standard tool in such studies.
This formalism parameterizes whatever new physics may appear in
theories that generalise the SM as effective operators of dimension
greater than four. Our goal is to provide the reader with as many
analytical expressions as possible for physical quantities which
might be measured at the ILC. In this way, our experimentalist
colleagues will have expressions they can include in their Monte
Carlo simulations. This paper is structured as follows. In
section~\ref{sec:eff} we present the effective operator formalism
and list the operators which contribute to lepton-violating
interactions, both interactions with gauge bosons and four-fermion
contact terms. In section~\ref{sec:lag} we use the existing
experimental bounds on decays such as $\mu\,\rightarrow\,e\,\gamma$
to exclude several of the operators which could {\em a priori} have
a contribution to the processes we will be considering. We also
analyse the role that the equations of motion of the fields play in
further simplifying our calculations. Having chosen a set of
effective operators, we proceed, in section~\ref{sec:dec}, to
calculate their impact on LFV decays of leptons, deducing analytical
expressions for those quantities. Likewise, in
section~\ref{sec:cross}, we will present analytical results for the
cross sections and asymmetries of several LFV processes which might
occur at the ILC. We analyse these results in section~\ref{sec:res},
performing a scan of a wide range of values for the anomalous
couplings we introduced, and considering their possible
observability at the ILC.

\section{Flavour changing effective operators}
\label{sec:eff}

\noindent The effective operator formalism of Buchm\"uller and
Wyler~\cite{buch} is based on the assumption that the Standard
Model (SM) of particle physics is the low energy limit of a more
general theory. Such theory would be valid at very high energies
but, at a lower energy scale $\Lambda$, we would only perceive its
effects through a set of effective operators of dimensions higher
than four. Those operators would obey the gauge symmetries of the
SM, and be suppressed by powers of $\Lambda$. This allows us to
write the effective lagrangian as a series, such that %%
\begin{equation}
{\cal L} \;\;=\;\; {\cal L}^{SM} \;+\; \frac{1}{\Lambda}\,{\cal
L}^{(5)} \;+\; \frac{1}{\Lambda^2}\,{\cal L}^{(6)} \;+\;
O\,\left(\frac{1}{\Lambda^3}\right) \;\;\; , \label{eq:l}
\end{equation}
where ${\cal L}^{SM}$ is the SM lagrangian and ${\cal L}^{(5)}$ and
${\cal L}^{(6)}$ contain all the dimension five and six operators
which, like ${\cal L}^{SM}$, are invariant under the gauge
symmetries of the SM. The ${\cal L}^{(5)}$ terms break baryon and
lepton numbers. Hence, we should start by considering the dimension
five LFV terms. However, these terms would also be responsible for
the generation of neutrino masses. With the present limits on
neutrino masses, $m_{\nu}<2$ eV~\cite{pdg}, the scale of new physics
would have to be of the order $10^{13}$ GeV~\cite{buch}, which is
clearly out of the reach of the next colliders.
This leaves us with the ${\cal L}^{(6)}$ operators, some of which,
after spontaneous symmetry breaking, generate dimension five terms.
The list of dimension six operators is quite vast~\cite{buch}. In
this work we are interested in those operators that give rise to
LFV. Throughout this paper we will use $l_h$ to represent a heavy
lepton and $l_l$ denotes a light one (whose mass we consider zero).
In processes where a tau lepton is present, both the muon and the
electron will be taken to be massless. If a given process only
involves muons and electrons, then the electron mass will be set to
zero, but the muon mass will be kept. Whenever the lepton's mass has
no bearing on the result we will use $l$ for all massless leptons,
and drop the generation index.

The effective operators that will be important for our studies fall
in three categories: (a) those that generate flavour-violating
vertices of the form $Z \,l_h\, l_l$ and $\gamma \,l_h \,l_l$ (and
also, for some operators, vertices like $\gamma\,\gamma \,l_h
\,l_l$); these operators always involve gauge fields, either
explicitly or in the form of covariant derivatives. (b) Four-fermion
operators, involving only leptonic spinors. (c) And a type of
operator that involves only scalar and fermionic fields that will
roughly correspond to a wave function renormalization of the fermion
fields.

\subsection{Effective operators generating  $Z \,l_h\, l_l$ and $\gamma \,l_h \,l_l$ vertices}
\noindent There are five tree-level dimension 6 effective operators
that can generate a new $Z \,l_h\, l_l$ interaction. This means that
these interactions are compatible with SM symmetries at tree level.
Following the notations of~\cite{buch} we write the first two
operators as
\begin{align}
{\cal O}_{D_e}
&=\frac{\eta^{R}_{ij}}{\Lambda^2}\,\left(\bar{\ell}^i_L\,
D^{\mu}\,e^j_R\right)\, D_{\mu} \phi \, \;\;\;,\;\;\;   {\cal
O}_{\bar{D}_e}=\frac{\eta^{L}_{ij}}{\Lambda^2}\,\left( D^{\mu}
\bar{\ell}^i_L\, \,e^j_R\right)\, D_{\mu} \phi  \;\;\; .
\label{eq:op1}
\end{align}
The coefficients $\eta^{R(L)}_{ij}$ are complex dimensionless
couplings and the $(i,j)$ are flavour indices. For flavour
violation to occur, these indices must differ. $\ell^i_L$ is a
left-handed $SU(2)$ doublet, $e^j_R$ is a right-handed $U(1)_{Y}$
singlet, $\phi$ is the Higgs scalar $SU(2)$ doublet. Notice that
the terms contributing to the interaction $Z \,l_h\, l_l$  in
which we are interested appear in the lagrangian when the Higgs
doublet acquires a vacuum expectation value (vev) $v$. There is no
$\gamma\, l_h\, l_l$ interaction stemming from these terms,
although one may obtain contributions to vertices involving also a
Higgs field, such as $\gamma\, \phi\, l_h\, l_l$ and $Z\, \phi\,
l_h\, l_l$.

The remaining three operators that contribute to the vertices $Z
\,l_h\, l_l$ but not to $\gamma, l_h\, l_l$ are given by
\begin{align}
{\cal O}_{\phi e} &= \;\;i\,
\;\;\frac{\theta^{R}_{ij}}{\Lambda^2}\,\left(\phi^{\dagger} D_{\mu}
\phi \right)\left(\bar{e}^i_R \, \gamma^{\mu} \,e^j_R \, \right)
\;\,\; , \nonumber \vspace{0.2cm} \\
{\cal O}^{(1)}_{\phi \ell} &= \;\;i\,
\;\;\frac{\theta^{L(1)}_{ij}}{\Lambda^2}\,\left(\phi^{\dagger}
D_{\mu} \phi \right)\left(\bar{\ell}^i_L \, \gamma^{\mu} \,\ell^j_L
\, \right) \;\,\; ,  \qquad \qquad {\cal O}^{(3)}_{\phi \ell} =
\;\;i\,
\;\;\frac{\theta^{L(3)}_{ij}}{\Lambda^2}\,\left(\phi^{\dagger}
D_{\mu} \tau_{I} \phi \right)\left(\bar{\ell}^i_L \, \gamma^{\mu}
\tau^{I} \,\ell^j_L \, \right) \;\;\; .
 \label{eq:op2}
\end{align}
Again, $\theta^{R}_{ij}$ and $\theta^{L(1),(3)}_{ij}$ are complex
dimensionless couplings, and the contributions to $Z \,l_h\, l_l$
arise when both scalar fields acquire a vev $v$. Because the
covariant derivatives act on those same fields and the SM Higgs has
no coupling to the photon, there are no contributions to $\gamma
\,l_h\, l_l$ from these operators. There are however five dimension
six operators that contribute to both the $Z \,l_h\, l_l$ and
$\gamma\, l_h\, l_l$ vertices and are only present at the one-loop
level. They are given by
\begin{align}
{\cal O}_{eB} &=
\;\;i\,\frac{\alpha^{B\,R}_{ij}}{\Lambda^2}\,\left(\bar{e}^i_R\,
\gamma^\mu\,D^\nu\,e^j_R\right)\,B_{\mu\nu} \;\;\; , \nonumber
\vspace{0.3cm} \\
{\cal O}_{\ell B} &=
\;\;i\,\frac{\alpha^{B\,L}_{ij}}{\Lambda^2}\,\left(\bar{\ell}^i_L\,
\gamma^\mu\,D^\nu\,\ell^j_L\right)\,B_{\mu\nu} \;\;\; , \qquad
\qquad {\cal O}_{\ell W} =
\;\;i\,\frac{\alpha^{W\,L}_{ij}}{\Lambda^2}\,\left(\bar{\ell}^i_L \,
\tau_{I} \,  \gamma^\mu\,D^\nu\,\ell^j_L\right)\,W^{I}_{\mu\nu}
\;\;\; , \vspace{0.3cm} \nonumber \\
{\cal O}_{eB\phi} &=
\;\;\frac{\beta^{B}_{ij}}{\Lambda^2}\,\left(\bar{\ell}^i_L\,
\sigma^{\mu\nu}\,e^j_R\right)\, \phi\,B_{\mu\nu} \;\;\; ,  \qquad
\qquad \qquad {\cal
O}_{eW\phi}=\frac{\beta^{W}_{ij}}{\Lambda^2}\,\left(\bar{\ell}^i_L\,
\, \tau_{I} \, \sigma^{\mu\nu}\,e^j_R\right)\,
\phi\,W^{I}_{\mu\nu}\;\;\; .\label{eq:op4}
\end{align}
$\alpha_{ij}$ and $\beta_{ij}$ are complex dimensionless couplings,
$B_{\mu\nu}$ and $W^{I}_{\mu\nu}$ are the usual $U(1)_Y$ and
$SU(2)_L$ field tensors, respectively. These tensors ``contain" both
the photon and Z boson fields, through the well-known Weinberg
rotation. Thus they contribute to both $Z \,l_h\, l_l$ and $\gamma\,
l_h\, l_l$ when we consider the partial derivative of $D^\mu$ in the
equations~\eqref{eq:op4} or when we replace the Higgs field $\phi$
by its vev $v$ in them. We will return to this point in
section~\ref{sec:gamma}.

\subsection{Four-Fermion effective operators producing an $e\,e\, l_h l_l$
contact interaction}
\noindent Because we are specifically interested in studying the
phenomenology of the ILC, we will only consider four-fermion
operators where two of the spinors involved correspond to the
colliding electrons/positrons of that collider. Another spinor will
correspond to a heavy lepton, $l_h$. There are four relevant types
of four-fermion operators that contribute to $e^+\,e^- \to l_h\,
l_l$,
\begin{eqnarray}
{\cal O}_{\ell \ell}^{(1)} &=& \frac{\kappa_{\ell \ell}^{(1)}}{2}
\left( \bar\ell_L \gamma_\mu \ell_L \right ) \left(\bar \ell_L
\gamma^\mu \ell_L \right)\;\;\; ,
\qquad \qquad {\cal O}_{\ell \ell}^{(3)} = \frac{\kappa_{\ell
\ell}^{(3)}}{2} \left( \bar\ell_L \gamma_\mu \tau^I \ell_L \right )
\left(\bar \ell_L \gamma^\mu \tau^I \ell_L \right)\;\;\;, \nonumber
\vspace{0.3cm} \\
{\cal O}_{e e} &=& \frac{\kappa_{e e}}{2} \left( \bar e_R \gamma_\mu
e_R \right ) \left(\bar e_R \gamma^\mu e_R \right) \;\;\; ,
\qquad \qquad {\cal O}_{\ell e} = \kappa_{\ell e} \left( \bar \ell_L
e_R \right ) \left(\bar e_R \ell_L \right) \;\;\;. \label{eq:ff}
\end{eqnarray}
Again, all of the couplings in these operators are, in general,
complex. As we have done with the previous operators, we should now
consider all possible ``placements" of the $l_h$ spinor, and
consider different couplings for each of them. But that would lead
to an unmanageable number of fermionic operators, all with the same
Lorentz structure but differing simply in the location of the heavy
lepton spinor. Thus we will simplify our approach and define only
one coupling constant for each type of operator. An exception is the
operator ${\cal O}_{\ell e} = \kappa_{\ell e} \left( \bar \ell_L e_R
\right ) \left(\bar e_R \ell_L \right)$, which corresponds to an
interaction between a right-handed current and a left-handed one.
Depending on where we place the $l_h$ spinor, then, we might have
two different effective operators. For example, if we consider the
operators that would contribute to $e^+\,e^- \to \tau^-\, e^+$, the
two possibilities we would have, putting the chiral structure of the
operators in evidence, are
\begin{equation}
{\cal O}_{\tau e} \;=\;\kappa_{\tau e}\,
\left(\bar{\ell}_L^\tau\,\gamma_R\,e_R\right) \left(\bar e_R
\,\gamma_L\, \ell^e_R \right) \;\;\; , \;\;\;{\cal O}_{e \tau}
\;=\;\kappa_{e \tau}\,\left(\bar{\ell}^e_L\,\gamma_R\,e_R\right)
\left(\bar \tau_R \,\gamma_L\, \ell^e_L \right) \;\;\;,
\end{equation}
where $\ell^e$ and $\ell^\tau$ are the leptonic doublets from the
first and third generations, respectively, and $\gamma_{L,R} = (1
\mp \gamma_5)/2$ are the usual chiral projectors. As we see, we find
two different Lorentz structures depending on where we ``insert" the
$\tau$ spinor. Therefore we define two different couplings, each
corresponding to the two possible flavour-violating interactions.

\noindent It will be simpler, however, to parameterize the
four-fermion effective Lagrangian built with the operators above in
the manner of ref.~\cite{Bar-Shalom:1999iy}. For the $e^+e^- l_h
l_l$ interaction, we have
\begin{equation}
{\cal L}_{ee l_l l_h}  =  {1\over\Lambda^2} \sum_{I,J=L,R} \biggl[
V_{IJ}^s \left({\bar e} \gamma_\mu \gamma_I e \right) \left( \bar
l_h \gamma^\mu \gamma_J l_l \right) + S_{IJ}^s \left( {\bar e}
\gamma_I e \right) \left( \bar l_h \gamma_J l_l \right)  \biggr]
\;\;\; .
\end{equation}
The vector-like ($V_{IJ}$) and scalar-like ($S_{IJ}$) couplings may
be expressed in terms of the coefficients of the four four-fermion
operators written in eq.~\eqref{eq:ff}~\cite{note1} in the following
manner:

\begin{eqnarray}
&& V_{LL}=\frac{1}{2} \left( \kappa_{\ell \ell}^{(1)} -
\kappa_{\ell \ell}^{(3)} \right)~,~ V_{RR}=\frac{1}{2} \kappa_{e
e}  ~, ~ V_{LR}=0~, ~
V_{RL}=0,\nonumber \\
&& S_{RR}=0 ~, ~ S_{LL}=0~,~S_{LR}=\kappa_{\ell e}^L
~,~S_{RL}=\kappa_{\ell e}^R~ .
\end{eqnarray}

\subsection{Effective operators generating an $l_h \,l_l$ mixing}
\noindent There is a special kind of interaction that corresponds to
a wave-function renormalization, which has its origin in the
operator
\begin{align}
{\cal O}_{e \ell \phi} &=
\;\;\frac{\delta_{ij}}{\Lambda^2}\,\left(\phi^{\dagger} \phi
\right)\left(\bar{\ell}^i_L\, \,e^j_R \, \phi \right) \;\; ,
\label{eq:delt}
\end{align}
where $\delta_{ij}$ are complex dimensionless couplings. After
spontaneous symmetry breaking the neutral component of the field
$\phi$ acquires a vev ($\phi_0\,\rightarrow\,\phi_0\,+\,v$, with
$v\,=\, 246/\sqrt{2}$ GeV) and a dimension three operator is
generated which is a flavour-violating self-energy like term. In
other words, it mixes, at the level of the propagator, the leptons
of different families. We consider these operators here for
completeness, even though we will show that they have no impact in
the phenomenology whatsoever.

\section{The complete Lagrangian}
\label{sec:lag}
\noindent The complete effective lagrangian can now be written as
a function of the operators defined in the previous section
\begin{align}
{\cal L}\;\; = &\;\; {\cal L}_{ee l_l l_h}\;+\;{\cal O}_{D_e}\;+\;
{\cal O}_{\bar{D}_e}\;+\; {\cal O}_{\phi e}\;+\; {\cal
O}^{(1)}_{\phi
\ell}\;+\; {\cal O}^{(3)}_{\phi \ell}\;+\; \nonumber \\
& \;\; {\cal O}_{e \ell \phi}\;+\;{\cal O}_{eB} \;+\;{\cal O}_{\ell
B}\;+\;{\cal O}_{\ell W}\;+\;{\cal O}_{eB\phi}\;+\;{\cal
O}_{eW\phi}\;+\; \mbox{h.c.} \, \, \, .
\end{align}
This lagrangian describes new vertices of the form $\gamma
\,\bar{l_h}\, l_l$, $Z \,\bar{l_h}\, l_l$, $\bar{e} \,e
\,\bar{l_h}\, l_l$, $\bar{l_h} l_l$ (and many others) and all of
their charge conjugate vertices. We will also consider an analogous
lagrangian with flavour indices exchanged - in other words, we will
consider couplings of the form $\eta_{hl}$ and $\eta_{lh}$, for
instance - except for the four fermion lagrangian, as was explained
in the previous section. Rather than write the Feynman rules for
these anomalous vertices and start the calculation of all LFV decay
widths and cross sections, we shall use all experimental and
theoretical constraints to reduce as much as possible the number of
independent couplings. After imposing these constraints we will
write the Feynman rules for the remaining lagrangian and proceed
with the calculation.

\subsection{The constraints from $l_h \rightarrow l_l \,\gamma$}
\label{sec:gamma}

\noindent Some of the operators presented in the previous section
can be immediately discarded due to the very stringent experimental
bounds which exist for the decays $\tau \rightarrow \mu\, \gamma$,
$\tau \rightarrow e\, \gamma$ and $\mu \rightarrow e\, \gamma$. The
argument is as follows: all the operators in eqs.~\eqref{eq:op4}
contribute to both $\gamma\,l_h\,l_l$ and $Z\,l_h\,l_l$
interactions, due to the presence of the gauge fields $B_\mu$ and
$W^3_\mu$ in the field tensors $B_{\mu\nu}$ and $W_{\mu\nu}$ that
compose them. Then we can write, for instance, an operator
$\cal{O}_{\ell \gamma}$, given by
\begin{equation}
{\cal O}_{\ell\gamma}\,=
\;\;i\,\frac{\alpha^{\gamma\,L}_{ij}}{\Lambda^2}\,\left(\bar{\ell}^i_L\,
\gamma^\mu\,\partial^\nu\,\ell^j_L\right)\,F_{\mu\nu}
\end{equation}
where $F_{\mu\nu}$ is the usual electromagnetic tensor. This
operator was constructed from both $\cal{O}_{\ell B}$ and ${\cal
O}_{\ell W}$, and the new effective coupling
$\alpha^{\gamma\,L}_{ij}$ is related to $\alpha^{B\,L}_{ij}$ and
$\alpha^{W\,L}_{ij}$ through the Weinberg angle $\theta_W$ by
\begin{equation}
\alpha^{\gamma\, L}_{i j}\;=\;\cos\theta_W \, \alpha^{B\,L}_{i j} -
\sin\theta_W \, \alpha^{W\, L}_{i j} \;\;\; . \label{eq:wein1}
\end{equation}
Following the same exact procedure we can also obtain an operator
${\cal O}_{\ell Z}$, with coupling constant given by
\begin{equation}
\alpha^{Z\, L}_{i j}\;=\;-\sin\theta_W \, \alpha^{B\,L}_{i j} -
\cos\theta_W \, \alpha^{W\, L}_{i j} \;\;\; .
\end{equation}
New operators with photon and Z interactions appear from the
remaining terms, with coupling constants given by
\begin{align}
\alpha^{\gamma\, R}_{i j}\;=\;\cos\theta_W \, \alpha^{B\,R}_{i j} &
& \alpha^{Z\, R}_{i j}\;=\;-\sin\theta_W \, \alpha^{B\,R}_{i j}
\nonumber \\
\beta^{\gamma\, R}_{i j}\;=\;\cos\theta_W \, \beta^{B}_{i j} & &
\beta^{Z\, R}_{i j}\;=\;-\sin\theta_W \, \beta^{B}_{i j} \nonumber \\
\beta^{\gamma\, L}_{i j}\;=\;\cos\theta_W \, \beta^{B}_{i j} -
\sin\theta_W \, \beta^{W}_{i j} & & \beta^{Z\, L}_{i
j}\;=\;-\sin\theta_W \, \beta^{B}_{i j} - \cos\theta_W \,
\beta^{W}_{i j} \;\;\; . \label{eq:wein3}
\end{align}
It is a simple matter to obtain the Feynman rules for the
$\gamma\,l_h\,l_l$ interactions from the lagrangian (they are
identical in form to those obtained for the flavour-violating
interactions $g\,t\,c$ and $g\,t\,u$ in
refs.~\cite{Ferreira:2005dr,Ferreira:2006xe}).
\begin{figure}[htbp]
  \begin{center}
    \epsfig{file=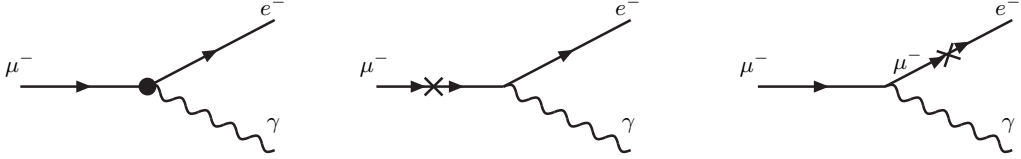,width=14 cm}
    \caption{$\mu \rightarrow e \gamma$ with effective anomalous vertices involving the couplings $\alpha$, $\beta$
and $\delta$.}
    \label{fig:decayegamma}
  \end{center}
\end{figure}
In figure~\ref{fig:decayegamma} we present the Feynman diagrams for
the decay $\mu \rightarrow e \gamma$ (in fact, for any decay of the
type $l_h \rightarrow l_l \gamma$) with vertices containing the
effective couplings $\alpha$, $\beta$ and $\delta$. Interestingly,
the $\delta$ contributions cancel out, already at the level of the
amplitude~\cite{note2}. The calculation of the remaining diagram is
quite simple and gives us the following expression for the width of
the anomalous decay $l_h \rightarrow l_l \gamma$ in terms of the
$\alpha$ and $\beta$ couplings:
\begin{eqnarray}
\Gamma(l_h\rightarrow l_l\gamma)&=&\frac{m_h^3}{64 \pi
\Lambda^4}\biggl[
 m_h^2(|\alpha^{\gamma R}_{lh}+\alpha^{{\gamma R}*}_{hl}|^2+|\alpha^{\gamma L}_{lh}+
 \alpha^{{\gamma L}*}_{hl}|^2) + 16v^2(|\beta^{\gamma}_{lh}|^2+|\beta^{\gamma}_{hl}|^2)\nonumber\\
&+&  8m_hv\: \mbox{Im}(\alpha^{\gamma
R}_{hl}\beta^{\gamma}_{hl}-\alpha^{\gamma R}_{lh}\beta^{\gamma
*}_{hl}-\alpha^{\gamma L}_{lh}\beta^{\gamma *}_{lh}+\alpha^{\gamma L}_{hl}\beta^{\gamma}_{lh})
\biggr] \, \, . \label{eq:wid}
\end{eqnarray}
So, for the decay $\mu \rightarrow e \gamma$, using the data
from~\cite{pdg}, we get (with $\Lambda$ expressed in TeV)
\begin{eqnarray}
\mbox{BR}(\mu\rightarrow e \gamma)&=& \frac{0.22}{\Lambda^4} \biggl[
(|\alpha^{\gamma R}_{e \mu}+\alpha^{{\gamma R}*}_{\mu
e}|^2+|\alpha^{\gamma L}_{e \mu}+\alpha^{{\gamma L}*}_{\mu e}|^2) +
4.3 \times 10^{7}(|\beta^{\gamma}_{e \mu}|^2+|\beta^{\gamma}_{\mu
e}|^2)\nonumber\\
&+&  1.3 \times 10^{4} \: \mbox{Im}(\alpha^{\gamma R}_{\mu
e}\beta^{\gamma}_{\mu e}-\alpha^{\gamma R}_{e \mu}\beta^{\gamma
*}_{\mu e}-\alpha^{\gamma L}_{e \mu}\beta^{\gamma *}_{e
\mu}+\alpha^{\gamma L}_{\mu e}\beta^{\gamma}_{e \mu}) \biggr]
\quad \, .
\end{eqnarray}
Now, all decays $l_h \rightarrow l_l \gamma$ are severely
constrained by experiment \cite{pdg} especially in the case of $\mu
\rightarrow e \gamma$ but also in $\tau \rightarrow e \gamma$ and
$\tau \rightarrow \mu \gamma$. To obtain a crude constraint on the
couplings, we can use the experimental constraint BR$(\mu
\rightarrow e \gamma) < 1.2 \times 10^{-11}$ and set all couplings
but one to zero. With this procedure we get the approximate bound
\begin{equation}
\frac{|\alpha_{e \mu}^{\gamma\,L,R}|}{\Lambda^2} \, \leqslant \, 7.4
\times 10^{-6}\;\;\; \mbox{TeV}^{-2}
\end{equation}
and identical bounds for the $\alpha_{\mu e}^{\gamma\,L,R}$
couplings. The constraints on the $\beta$ constants are roughly four
orders of magnitude smaller. Using the same procedure for the two
remaining LFV processes we get
\begin{equation}
\frac{|\alpha_{e \tau}^{\gamma\,L,R}|}{\Lambda^2} \, \leqslant \,
1.6\times 10^{-3} \;\;\; \mbox{TeV}^{-2}\qquad \quad
\frac{|\alpha_{\mu \tau}^{\gamma\,L,R}|}{\Lambda^2} \, \leqslant \,
1.3 \times 10^{-3} \;\;\; \mbox{TeV}^{-2}
\end{equation}
with the $\beta$ couplings even more constrained in their values.

The experimental bounds on the various branching ratios are so
stringent that they pretty much curtail any possibility of these
anomalous operators having observable effects on any experiences
performed at the ILC. To see this, let us consider the
flavour-violating reaction $\gamma\,\gamma\,\rightarrow\,l_h\,l_l$,
which in principle could occur at the ILC~\cite{ggilc}. There are
five Feynman diagrams involving the $\{\alpha\,,\,\beta\}$ couplings
that contribute to this process.
%They are similar to the diagrams
%shown in figure~2 of ref.~\cite{Ferreira:2006xe} for the reaction
%$g\,g\,\rightarrow\,\bar{t}\,u$, if one excludes the diagram with
%the triple gluon vertex.
There are also three diagrams involving the
$\delta$ couplings of eq.~\eqref{eq:delt}, but their contributions
(once again) cancel at the level of the amplitude. The calculation
of the cross section for this process is laborious but unremarkable.
The end result, however, is extremely interesting. The cross section
is found to be
\begin{equation}
\frac{d\sigma(\gamma\,\gamma\,\rightarrow \,l_h\,l_l)}{d t} \;
\;=\;\; -\,\frac{4\pi\alpha\,F_{\gamma\gamma}}{{m_h}^3\,s\,
    {( {m_h}^2 - t ) }^2\,t\,{( {m_h}^2 - u ) }^2\,u}\;\Gamma(l_h\,\rightarrow
    \,l_l\,\gamma)\;\;\;,
\label{eq:ggl}
\end{equation}
with a function $F_{\gamma\gamma}$ given by
\begin{align}
F_{\gamma\gamma} &=\;\;\;
      m_h^{10}\,( t + u )- 12 \,m_h^8\,t\,u +
      m_h^6\,( t + u )\,( t^2 + 13\,t\,u + u^2 )  - m_h^4\,t\,u\,( 7\,t^2 + 24\,t\,u + 7\,u^2 )\nonumber \\
      & \;\;\; \;\;\;+\,12\,m_h^2\,t^2\,u^2\,( t + u ) - 6\,t^3\,u^3 \;\;\; .
      \label{eq:fgg}
\end{align}
The remarkable thing about eq.~\eqref{eq:ggl} is the
proportionality of the (differential) cross section to the width
of the anomalous decay $l_h\,\rightarrow\,l_l\,\gamma$, which is
to say (modulus the total width of $l_h$, which is well known), to
its branching ratio. A similar result had been obtained for
gluonic flavour-changing vertices in
refs.~\cite{Ferreira:2005dr,Ferreira:2006xe}. Because the allowed
branching ratios for the $l_h\,\rightarrow\,l_l\,\gamma$ are so
constrained, the predicted cross sections for the ILC are
extremely small. We have
\begin{align}
\sigma(\gamma\,\gamma\,\rightarrow \,\mu^-\,e^+)\;\sim&
\;\;\;10^{-8}\,\times\,\mbox{BR}(\mu\,\rightarrow \,e\,\gamma) \;\;\mbox{pb} \nonumber \\
\sigma(\gamma\,\gamma\,\rightarrow \,\tau^-\,\mu^+)\;\sim&
\;\;\;10^{-5}\,\times\,\mbox{BR}(\tau\,\rightarrow \,\mu\,\gamma) \;\;\mbox{pb}  \nonumber \\
\sigma(\gamma\,\gamma\,\rightarrow \,\tau^-\,e^+)\;\sim&
\;\;\;10^{-5}\,\times\,\mbox{BR}(\tau\,\rightarrow \,e\,\gamma)
\;\;\mbox{pb} \;\;\; , \label{eq:crl}
\end{align}
with $\sqrt{s}\,=\,1$ TeV. With the current branching ratios of the
order of $10^{-12}$ for the muon decay and $10^{-7}$ for the tau
ones, it becomes obvious that these reactions would have
unobservable cross sections.

Our conclusion is thus that the $\alpha_{ij}^{\gamma}$ and
$\beta_{ij}^{\gamma}$ couplings are too small to produce observable
signals in foreseeable collider experiments. However, both
$\{\alpha^{\gamma}_{i j}\,,\,\beta^{\gamma}_{i j}\}$ and
$\{\alpha^{Z}_{i j}\,,\,\beta^{Z}_{i j}\}$ are written in terms of
the original $\{\alpha^{B,W}_{i j}\,,\,\beta^{B,W}_{i j}\}$
couplings, via coefficients (sine and cosine of $\theta_W$) of order
1. Hence, unless there was some bizarre unnatural cancellation, the
couplings $\{\gamma\,,\,Z\}$ and $\{B\,,\,W\}$ should be of the same
order of magnitude. Since we have no reason to assume such a
cancellation, we come to the conclusion that the $\alpha$ and
$\beta$ couplings are simply too small to be considered interesting.
They will have no bearing whatsoever on anomalous LFV interactions
mediated by the $Z$ boson. From now on, we will simply consider them
to be zero, which means that there will not be any anomalous
vertices of the form $\gamma\,l_i\,l_j$.
\subsection{A set of free parameters}
\begin{figure}[htbp]
  \begin{center}
    \epsfig{file=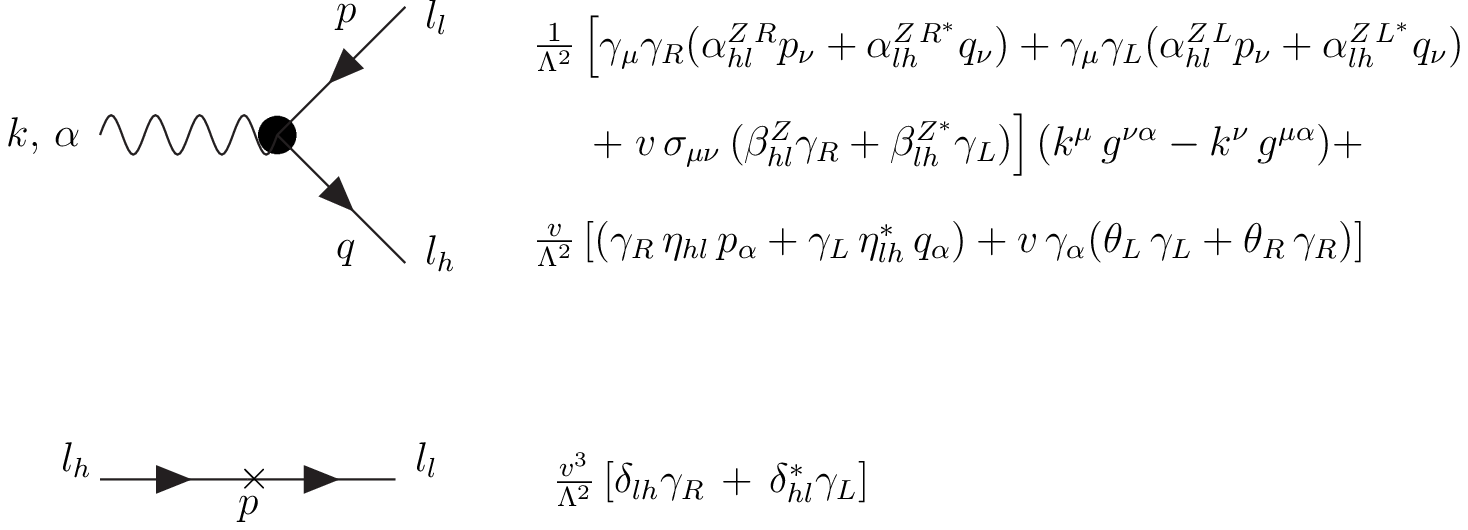,width=14 cm}
    \caption{Feynman rules for anomalous $Z \, \bar{l_h} \, l_l$ and
$\bar{l_l} \, l_h$ vertices.}
    \label{fig:feynman}
  \end{center}
\end{figure}

\noindent In the previous section we have presented the complete
set of operators that give contributions to the flavour violating
processes $e^+ e^- \rightarrow l_h l_l$. However, these operators
are not, {\em a priori}, all independent. It can be shown that
(see refs.~\cite{buch,Ferreira:2005dr,Grzadkowski:2003tf,Ferr3}
for details), for instance, there is a relation between operators
of the types ${\cal O}_{eB\phi}$ and ${\cal O}_{eB}$ and some of
the four fermion operators, modulo a total derivative. These
relations between operators appear when one uses the fermionic
equations of motion, along with integration by parts. They could
be used to discard operators whose coupling constants are $\alpha$
and $\beta$, or some of the four fermion operators. We used this
argument to present the results in
refs.~\cite{Ferreira:2005dr,Ferreira:2006xe,Ferr3} in a more
simplified fashion. However, in the present circumstances, we
already discarded the $\alpha$ and $\beta$ operators due to the
size of their contributions to physical processes being extremely
limited by the existing bounds on flavour-violating leptonic
decays with a photon. Since we already threw away these two sets
of operators, we are not entitled to use the equations of motion
to attempt to eliminate another.
\\
Notice also that in most of the work that was done with the
effective lagrangian approach one replaces, at the level of the
amplitude, operators of the type ${\cal O}_{D_e}$ by operators of
the type ${\cal O}_{eZ\phi}$ by using Gordon identities. In fact, it
can be shown that the following relation holds for free fermionic
fields,
\begin{equation}
\bar{e}^i_L\, \partial^{\mu}\,e^j_R\,  = \, m_j \, \bar{e}^i_L\,
\gamma^{\mu} \,e^j_R - \bar{e}^i_L\, \sigma^{\mu\alpha}\
\partial^{\alpha}\,e^j_R \, \, .
\end{equation}
Notice that the use of Gordon identities is not the same thing as
using the field's equations of motion to eliminate operators: in the
latter case, one proves that different operators are related to one
another and use those conditions to choose among them; in the
former, all we are doing is re-writing the amplitude in a different
form. And in our case, this procedure does not bring any
simplification.
\\
Finally, using the equations of motion, a relation can be
established between operators ${\cal O}_{D_e}$ and ${\cal
O}_{\bar{D}_e}$, namely
\begin{equation}
{\cal O}_{D_e} + {\cal O}_{\bar{D}_e} + \left(\bar{\ell}_L \,e_R
\right) \, \left[ \Gamma_e^{\dagger} \bar{e}_R \,\ell_L + \Gamma_u
\bar{q}_L \epsilon \,u_R  +  \Gamma_d^{\dagger} \bar{d}_R \, q_L
\right] \, = \, 0 \;\;\;,
\end{equation}
where the $\Gamma_e$ coefficients are the leptonic Yukawa couplings
and $\epsilon$ the bidimensional Levi-Civita tensor. We see that the
relationship between these two operators involves four-fermion terms
as well. This relation means we can choose between one of the two
operators ${\cal O}_{D_e}$ and ${\cal O}_{\bar{D}_e}$, given that
the four-fermion operators appearing in this expression have already
been considered by us. This means that only one of the $\eta_{ij}^R$
and $\eta_{ij}^L$ couplings will appear in the calculation. We chose
the first one and will drop, from this point onwards, the
superscript ``$R$". Also, after expanding the operators of
eq.~\eqref{eq:op2}, we see that the $\theta$ couplings always appear
in the same combinations. We therefore define two new couplings,
$\theta_R$ and $\theta_L$, as
\begin{equation}
\theta_R \; = \; \theta_{lh}^R + \theta_{hl}^{R*}\;\;\; , \;\;\;
\theta_L \; = \; \theta_{lh}^{L(1)} +
\theta_{hl}^{L(1)*}-\theta_{lh}^{L(3)} - \theta_{hl}^{L(3)*}\;\;\;
 .
\end{equation}
As an aside, we must add that the use of equations of motion to
simplify the effective lagrangian is not followed by all authors.
For instance, the authors of~\cite{whi} do not use them and consider
instead a fully general set of dimension six effective operators.
\\
The original lagrangian is now reduced to
\begin{equation}
{\cal L}\;\; =\;\;\; {\cal L}_{ee l_l l_h}\;+\;{\cal O}_{D_e}\;+\;
{\cal O}_{\phi e}\;+\; {\cal O}^{(1)}_{\phi \ell}\;+\; {\cal
O}^{(3)}_{\phi \ell}\;+\;{\cal O}_{e \ell \phi}\;+\; \mbox{h.c.}
\qquad .
 \label{eq:li}
\end{equation}
We will not present the Feynman rules for the four fermion
interactions because they are obvious and rather cumbersome to
write. The remaining Feynman rules we will use are presented in
figure~\ref{fig:feynman}, where lepton momenta follow the arrows and
vector boson momentum is incoming. For completeness, we included the
$\alpha^Z_{ij}$ and $\beta^Z_{ij}$ in this figure, but we remind the
reader that we have set them to zero.

\section{Decay widths}
\label{sec:dec}
\noindent As we said before, all LFV processes are severely
constrained by experimental data. Now that we have settled on a set
of anomalous effective operators, we should first consider what is
the effect of those operators on LFV decays. The existing data
severely constrains two types of decay: a heavy lepton decaying into
three light ones, $l_h\,\rightarrow\,l\,l\,l$, such as
$\tau^-\,\rightarrow\,e^-\,e^+\,e^-$, and decays of the Z boson to
two different leptons, $Z\,\rightarrow\,l_h\,l_l$ (such as
$Z\,\rightarrow\,\tau^+\,e^-$). Flavour-violating processes
involving neutrinos in the final state (such as, say,
$Z\,\rightarrow\,\nu_\tau\,\bar{\nu}_e$) are not constrained by
experimental data, as they are indistinguishable from the ``normal"
processes.

\begin{figure}[htbp]
  \begin{center}
    \epsfig{file=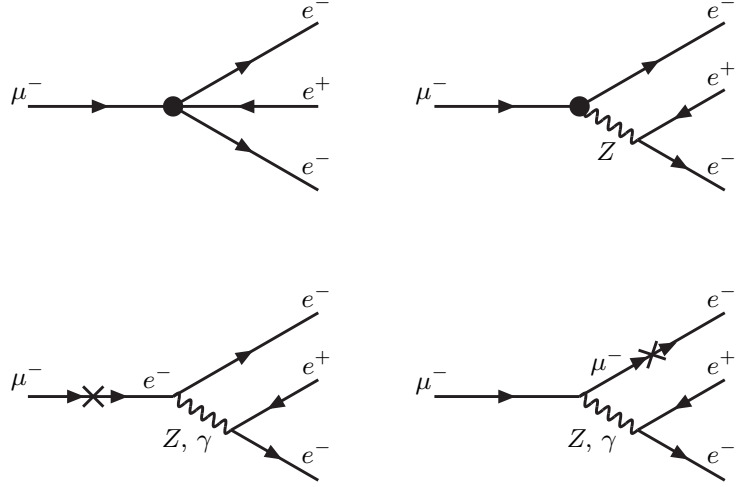,width=10 cm}
    \caption{Feynman diagrams for the decay $\mu^- \,\rightarrow \,e^- \,e^+ \,e^-$.}
    \label{fig:decayZ}
  \end{center}
\end{figure}

For the 3-lepton decay, there are three distinct contributions,
whose Feynman diagrams are shown in figure~\ref{fig:decayZ} for
the particular case of $\mu^- \,\rightarrow \,e^- \,e^+ \,e^-$. As
before, the contributions involving the $\delta$ operators cancel
at the level of the amplitude and have absolutely no effect on the
physics. Using the Feynman rules in figure~\ref{fig:feynman} and
the four fermion lagrangian we can determine the expression for
the decay $l_h \rightarrow l l l$. Remember that $l$ stands for a
massless lepton whatever its flavour is.
The decay width obtained is the sum of three terms, to wit
\begin{equation}
\Gamma (l_h \,\rightarrow\, l\, l\, l)\;=\;\Gamma_{4f}(l_h
\,\rightarrow\, l\, l\, l)\,+\,\Gamma_Z(l_h \,\rightarrow\, l\, l\,
l)\,+\,\Gamma_{int}(l_h \,\rightarrow\, l\, l\, l)\;\;\; ,
\end{equation}
where $\Gamma_{4f}$ contains the contributions from the
four-fermion graph in figure~\ref{fig:decayZ}, $\Gamma_Z$ those
from the Feynman diagram with a Z boson and $\Gamma_{int}$ the
interference between both diagrams. A simple calculation yields
\begin{align}
\Gamma_{4f} (l_h \,\rightarrow\, l\, l\, l) \;=&  \frac{m_h^5}{6144
\, \pi^3\, \Lambda^4}
\,\left[|S_{LR}|^2+|S_{RL}|^2+4(|V_{LL}|^2+|V_{RR}|^2)\right]
\nonumber \vspace{0.3cm} \\
\Gamma_Z (l_h \,\rightarrow\, l\, l\, l)     \;=&\;\;
\frac{(g_A^2+g_V^2) v^2}{768\, M_z^4 \,\pi^3 \,\Lambda^4} \, \left\{
\left( |\theta_L|^2+|\theta_R|^2 \right) v^2 m_h^5 + \frac{1}{2}
\mbox{Re} \left(\eta_{lh} \theta_L^* +  \eta_{hl} \theta_R^* \right)
v\, m_h^6 \right.
\nonumber \vspace{0.3cm} \\
& \;\; \left.  + \frac{m_h^7}{10 M_z^2} \left[ (|\eta_{lh}|^2  +
|\eta_{hl}|^2) M_z^2 + 6 (|\theta_L|^2  +  |\theta_R|^2)\, v^2
\right] \right\} \nonumber \\
\Gamma_{int} (l_h \,\rightarrow\, l\, l\, l)     \;=&\;\; \frac{v^2
\, m_h^5}{768 M_z^2 \pi^3 \Lambda^4} \,  \left[ \left( 1+ \frac{3
m_h^2}{10 M_z^2} \right) \biggl\{  \left( g_V +g_A \right) \mbox{Re}
\left( \theta_L V_{LL}^* \right) +  \right.
\nonumber \vspace{0.3cm} \\
& \;   \left( g_V -g_A \right) \mbox{Re} \left( \theta_R V_{RR}^*
\right) \biggr\}  - \frac{m_h}{4v} \left( 1+ \frac{m_h^2}{5 M_z^2}
\right) \biggl\{ \left( g_V +g_A \right) \mbox{Re}
\left( \eta_{lh} V_{RR}^*  \right) +  \nonumber \vspace{0.3cm} \\
& \; \left( g_V -g_A \right) \mbox{Re} \left( \eta_{hl} V_{LL}
\right) \biggr\} \biggr]  \;\;\; . \label{eq:expa}
\end{align}
where \begin{equation} g_V \;=\;
-\,\frac{e}{\sin\theta_W\,\cos\theta_W}\left(-\frac{1}{4} +
\sin\theta_W^2\right)\;\;\; , \;\;\; g_A \;=\;
\,\frac{e}{4\sin\theta_W\,\cos\theta_W} \end{equation}
and $e$ is the elementary electric charge. An important remark about
these results: they are not, in fact, the {\em exact} expressions
for the decay widths. The full expressions for $\Gamma_Z (l_h
\,\rightarrow\, l\, l\, l)$ and $\Gamma_{int} (l_h \,\rightarrow\,
l\, l\, l)$ are actually the sum of a logarithmic term and a
polynomial one. However, it so happens that the first four terms of
the Taylor expansion in $m_h/M_z$ of the logarithm cancel the
polynomial exactly. The expressions of eq.~\eqref{eq:expa} are
therefore the first surviving terms of that Taylor expansion, and
constitute an excellent approximation to the exact result, and one
that is much easier to deal with numerically (the cancellation
mentioned poses a real problem in numerical calculations).

As for the LFV decays of the Z-boson, there is an extensive
literature on this subject~\cite{zes}. There are, of course, no four
fermion contributions to this decay width, and a simple calculation
provides us the following expression:
\begin{align}
\Gamma (Z \,\rightarrow\, l_h\, l_l)     \;=&\;\;
\frac{(M_z^2-m_h^2)^2 \,v^2}{128\, M_z^5 \,\pi \,\Lambda^4} \,
\left[ (M_z^2-m_h^2)^2 (|\eta_{hl}|^2+|\eta_{lh}|^2) + 4\, (m_h^2+ 2
M_z^2) v^2 (|\theta_{L}|^2+|\theta_{R}|^2) \right.
\nonumber \vspace{0.3cm} \\
& \;  \left. \; + \,4\, m_h\, (m_h^2-M_z^2)\, v \,\mbox{Re} \left(
\theta_L \eta_{hl}+ \theta_R \eta_{lh}^* \right) \right] \;\;\; .
\end{align}

\section{Cross Sections}
\label{sec:cross}
\begin{figure}[htbp]
  \begin{center}
    \epsfig{file=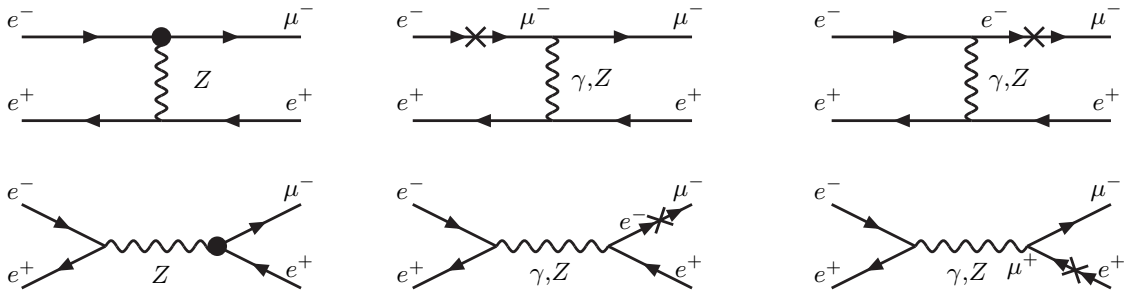,width=15 cm}
    \caption{Feynman diagrams describing the process $e^+ e^- \rightarrow \mu^- e^+$}
    \label{fig:cross}
  \end{center}
\end{figure}

\begin{figure}[htbp]
  \begin{center}
    \epsfig{file=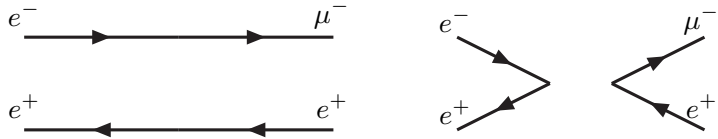,width=9.5 cm}
    \caption{Interpretation of the four fermion terms
             contributing to the process $e^+ e^- \rightarrow \mu^- e^+$
             in terms of currents; notice the analog of a $t$ channel
             and an $s$ one.}
    \label{fig:cross4f}
  \end{center}
\end{figure}

\noindent In this section we will present expressions for the cross
sections of various LFV processes that may occur at the ILC. There
are three such processes, namely (1) $e^+ e^- \rightarrow \mu^-
e^+$, (2) $e^+ e^- \rightarrow \tau^- e^+$ and (3) $e^+ e^-
\rightarrow \tau^- \mu^+$, as well as the respective
charge-conjugates. We have calculated all cross sections keeping
both final state masses. However, given the energies involved, the
contributions to the cross sections which arise from the lepton
masses are extremely small, and setting them to zero is an excellent
approximation. We thus present all formulae with zero leptonic
masses, as they are much simpler than the complete expressions. In
figures~\ref{fig:cross} and~\ref{fig:cross4f} we present all
diagrams that contribute to the process $e^+ e^- \rightarrow \mu^-
e^+$.
A brief word about our conventions. There are two types of LFV
production cross sections, corresponding to different sets of
Feynman diagrams. In the case of process (1), we see from
figures~\ref{fig:cross} and~\ref{fig:cross4f} that the reaction
can proceed through both a $t$-channel and an $s$-channel - this
is obvious for the diagrams involving the exchange of a photon or
a $Z$ boson. For the four-fermion channels less so, but
figure~\ref{fig:cross4f} illustrates the $t$ and $s$-channel
analogy. Depending on the ``location" of the incoming electron
spinor in the operators of eq.~\eqref{eq:ff}, we can interpret
those operators as two fermionic currents interacting with one
another, that interaction is obviously analog to the two different
channels. Process (2) has diagrams identical to those of process
(1). Process (3), however, can only occur through the $s$ channel
- that is obvious once one realizes that for process (3) there is
no positron in the final state. In fact for process (3) there are
only ``s-channel" contributions from the four-fermion operators.

A simple way of condensing the different four-fermion cross
sections into a single expression is to adopt the following
convention: we will include indices ``s" and ``t" in the four
fermion couplings. If we are interested in the cross sections for
processes (1) and (2) - which occur through both $s$ and $t$
channels - then all ``s" couplings will be equal to the ``t" ones.
If we wish to obtain the cross section for process (3) (which only
has $s$ channels) we must simply set all couplings with a ``t"
index to zero. We have further considered the likely possibility
that in the ILC one may be able to polarize the beams of incoming
electrons and positrons~\cite{Moortgat-Pick:2005cw}. Thus,
$\sigma_{IJ}$ represents the polarized cross section for an $I$
polarized electron and a $J$ polarized positron, with
$\{I\,,\,J\}\,=\,\{R\,,\,L\}$ - that is, beams with a right-handed
polarization or a left-handed one. The explicit expressions for
the four-fermion differential cross sections are then given by
\begin{align}
\frac{d \sigma_{LL}}{dt} \, = & \, \frac{1}{16 \pi s^2 \Lambda^4}
\left[ 4 u^2 |V_{LL}^s + V_{LL}^t|^2 + t^2 (4|V_{LR}^s|^2 +
|S_{RL}^t|^2)\right]
\nonumber \\
\frac{d \sigma_{RR}}{dt} \, = & \, \frac{1}{16 \pi s^2 \Lambda^4}
\left[ 4
u^2 |V_{RR}^s + V_{RR}^t|^2 +  t^2 (4|V_{RL}^s|^2 + |S_{LR}^t|^2) \right] \nonumber \\
\frac{d \sigma_{LR}}{dt} \, = & \, \frac{1}{16 \pi s^2 \Lambda^4}
\left[ u^2 |S_{LL}^s + S_{LL}^t|^2 + s^2 (4|V_{RL}^t|^2 +
|S_{LR}^s|^2) \right]
\nonumber \\
\frac{d \sigma_{RL}}{dt} \, = & \, \frac{1}{16 \pi s^2 \Lambda^4}
\left[ u^2 |S_{RR}^s + S_{RR}^t|^2 + s^2 (4|V_{LR}^t|^2 +
|S_{RL}^s|^2) \right]
\label{eq:sig4f}
\end{align}
See appendix~\ref{sec:cross2} for the full calculation. The
unpolarized cross section is obviously the averaged sum over the
four terms of eq.~\eqref{eq:sig4f}. To re-emphasize, the
four-fermion cross section for processes (1) and (2) is obtained
from this expression by setting all ``s" couplings equal to the
``t" ones; and to obtain the cross section for process (3) one
must simply set all ``t" couplings to zero.
\\
The total cross sections for each of the processes are then given by
\begin{align}
\sigma^{(1,2)}(e^- e^+ \rightarrow l_h e^+) &=\;\; \sigma_Z^{(1,2)}
\,+\, \sigma_{4f}^{(1,2)}\,+\, \sigma_{int}^{(1,2)}
\nonumber \\
\sigma^{(3)}(e^- e^+ \rightarrow \tau^- \mu^+) &=\;\; \sigma_Z^{(3)}
\,+\, \sigma_{4f}^{(3)} \,+\, \sigma_{int}^{(3)}\;\;\; ,
\end{align}
where $\sigma_Z$ is the cross section involving only the anomalous
$Z$ interactions of figure~\ref{fig:feynman}, $\sigma_{4f}$ the
four-fermion cross section - whose calculation we already
explained - and $\sigma_{int}$ the interference between both of
these. The $\delta$ couplings also present in
figure~\ref{fig:feynman} end up not contributing at all to the
physical cross sections, once again. For completeness, then, the
remaining terms in the differential cross section for processes
(1) and (2) are given by
\begin{equation}
\frac{d\sigma_{Z}^{(1,2)}}{dt}\; = - v^2 \, \frac{v^2 \, \left[F_1
(g_A,g_V) \, |\theta_L|^2 + F_1 (g_A,-g_V) |\theta_R|^2 \right] \, +
\, F_2 (g_A,g_V) \, |\eta_{lh}|^2 + F_2 (g_A,-g_V) |\eta_{hl}|^2
}{32\,\pi \,\Lambda^4\, {\left( M_z^2 - s \right) }^2\,s^2\,{\left(
M_z^2 - t \right) }^2}
\end{equation}
with
\begin{align}
F_1 (g_A,g_V) \,= & \, 2\,\left\{( g_A + g_V)^2 \left[ s\, t \, (2
\, M_z^4+2\, u \, M_z^2+s^2+t^2) - u \, M_z^2 \, (u\, M_z^2+2\,
s^2+2 \, t^2) \right]\right. +
\nonumber \\
& \left. 2 \, ( g_A^2 + g_V^2) u \, (t^3+s^3) + (g_A - g_V)^2 s \,
u^2 \, t \right\}
\nonumber \\
&
\nonumber \\
F_2 (g_A,g_V) \,= & \, - t \, u \, s \, \left[ ( g_A^2 + g_V^2)(3 \,
M_z^4+3\, u \, M_z^2+s^2+t^2+ s \, t) +2 \, g_A \, g_V \, \left(
M_z^2 - s \right) \,\left( M_z^2 - t \right) \right]\;\;\; .
\end{align}
The interference term is  given by
\begin{align}
\frac{d\sigma_{int}^{(1,2)}}{dt} \; = & \, \frac{(t-s) \, v^2
}{16\,\pi \,\Lambda^4\,
   \left( M_z^2 - s \right) \,s^2\,\left( M_z^2 - t \right)}
\nonumber \\
& \biggl[  \left(  g_A \,
         \mbox{Re} \left( \theta_L S_{LR}^* - \theta_R S_{RL}^* \right)   +
          g_V \, \mbox{Re} \left( \theta_L S_{LR}^* + \theta_R
S_{RL}^* \right)    \right) \,
        \left( s\,t + \left( - M_z^2 + s + t \right) \,u \right)
\nonumber \\
&       +\,4\, ( g_A - g_V) \,(s+t) \,u \, \mbox{Re} \left( \theta_L
V_{LL}^* \right) -
     4\,( g_A + g_V) \,(s+t) \,u  \, \mbox{Re} \left( \theta_R
V_{RR}^* \right) \biggr]
\end{align}
For process (3), we have
\begin{align}
\frac{d\sigma_{Z}^{(3)}}{dt} \,  = & \, \frac{- v^2 }{32\,\pi
\,\Lambda^4\,
   {\left(  M_z^2 - s \right) }^2\,s^2} \biggl\{ 2\, v^2 \left[
   (g_A^2+g_V^2)\, (2tu-s^2) \, (|\theta_L|^2+|\theta_R|^2) \right.
\nonumber \\
& \left. + \, 2 \, g_A \, g_V \,s (t-u)\, (|\theta_L|^2-
|\theta_R|^2) \right] - (g_A^2+g_V^2) \, t \, u \, s \,
(|\eta_{hl}|^2+|\eta_{lh}|^2) \biggr\}
\end{align}
and finally, the interference terms are
\begin{equation}
\frac{d\sigma_{int}^{(3)}}{dt}=\frac{(s+t) \, u \, v^2 }{8\,\pi
\,\Lambda^4\, \left( M_z^2 - s \right) \,s^2} \biggl[ ( g_V - g_A)
\, Re \left( \theta_L V_{LL}^* \right) +
   ( g_V + g_A)  \, \mbox{Re} \left( \theta_R
V_{RR}^* \right) \biggr]\;\;\; .
\label{eq:int}
\end{equation}
At this point we must remark on the different energy behavior that
these various terms follow. Once integrated in $t$, the
four-fermion terms grow linearly with $s$, whereas those arising
from the anomalous $Z$ couplings have a much smoother evolution
with $s$ - whereas the first ones diverge as
$s\,\rightarrow\,\infty$, the second ones tend to zero. See
appendix~\ref{sec:cross2} for the expressions of the integrated
cross sections. This could be interpreted as a clear dominance of
the four-fermion terms over the remaining anomalous couplings.
However, we must remember that we are working in a
non-renormalizable formalism. We know, from the beginning, that
these operators only offer a reasonable description of high-energy
physics up to a given scale, of the order of $\Lambda$. The
dominance of the four-fermion cross section must therefore be
carefully considered - it may simply happen, as there is nothing
preventing it, that the four-fermion couplings of
eq.~\eqref{eq:ff} are much smaller in size than the $Z$ boson ones
of figure~\ref{fig:feynman}.

As we saw, the $\delta$ couplings end up not contributing to
either decay widths or cross sections (and this is true regardless
of whether the light leptons are considered massless or not). As
we mentioned before, their inclusion could be interpreted as an
on-shell renormalization of the leptonic propagators. On that
light, their cancellation suggests that the effective operator
formalism is equivalent to an on-shell renormalization scheme.
This is further supported by the fact that the list of effective
operators of ref.~\cite{buch} was obtained by using the fields'
equations of motion to simplify several terms. However, we must
mention that at least in some Feynman diagrams (some of those
contributing to $\gamma\,\gamma\,\rightarrow\,l_h\,l_l$, for
instance), the ``$\delta$-insertions" were made in {\em internal}
fermionic lines, so that this cancellation is not altogether
obvious.

\subsection{Asymmetries}
\noindent In a collider with polarized beams, asymmetries can play a
major role in the determination of flavour-violating couplings. A
great advantage of using these observables is that, as will soon
become obvious, all dependence on the scale of unknown physics,
$\Lambda$, vanishes due to their definition. There is a strong
possibility that the ILC could have both beams polarized, therefore
the measurement of polarization asymmetries could be very
interesting. For a more detailed study
see~\cite{Moortgat-Pick:2005cw}. A particulary appealing situation
is found when the contributions from the $Z$ boson anomalous
couplings are not significant when compared with the four fermion
ones. In this case the study of asymmetries would allow us, in
principle, to determine each four-fermion coupling individually. We
will now concentrate on one of the most feasible scenarios, which is
to have a polarized electron beam and an unpolarized positron beam.
We will take both the right-handed and left-handed polarizations to
be 100\%, which is obviously above what is expected to occur (recent
studies show that a 90 \% polarization is
attainable)~\cite{Moortgat-Pick:2005cw}. The differential cross
sections for left-handed ($P_{e^-}=-1$) and right-handed
($P_{e^-}=+1$) polarized electrons are
\begin{align}
\frac{d \sigma_{L}}{dt} \, = & \, \frac{1}{16 \pi s^2 \Lambda^4}
\left( 4 u^2 |V_{LL}^s + V_{LL}^t|^2 + t^2 |S_{RL}^t|^2 + s^2
|S_{LR}^s|^2 \right)
\nonumber \\
\frac{d \sigma_{R}}{dt} \, = & \, \frac{1}{16 \pi s^2 \Lambda^4}
\left( 4 u^2 |V_{RR}^s + V_{RR}^t|^2 +  t^2 |S_{LR}^t|^2 + s^2
|S_{RL}^s|^2\right) \, \, .
\end{align}
Two forward-backward asymmetries for the left-handed and
right-handed polarized cross sections can now be defined as
\begin{equation}
A_{FB,L (R)} \, = \, \frac{\int_{0}^{\pi/2} d \sigma_{L (R)}
(\theta) \, - \, \int_{\pi/2}^{\pi} d \sigma_{L
(R)}(\theta)}{\sigma_{L (R)}}
\end{equation}
and we can also define a left-right asymmetry, given by
\begin{equation}
A_{LR} \, = \, \frac{\sigma_L -
\sigma_R}{\sigma_{L}+\sigma_{R}}\;\;\; ,
\end{equation}
where $\sigma_{L (R)}$ is the total cross section for a
left-handed (right-handed) polarized electron beam. Note that we
have assumed that the polarization of the final state particles is
not measured. Otherwise we could get even more information by
building an asymmetry related to the measured final state
polarizations. Using the expressions on appendix~\ref{sec:cross2}
it is simple to find, for these asymmetries, the following
expressions:
\begin{equation}
A_{FB,L} \, = \, \frac{12 \, |V_{LL}^s + V_{LL}^t|^2 - 3 \,
|S_{RL}^t|^2}{16 \, |V_{LL}^s + V_{LL}^t|^2 + 4 \, |S_{RL}^t|^2+ 12
\, |S_{LR}^s|^2}
\label{eq:afbl}
\end{equation}
and
\begin{equation}
A_{FB,R} \, = \, \frac{12 \, |V_{RR}^s + V_{RR}^t|^2 - 3 \,
|S_{LR}^t|^2}{16 \, |V_{RR}^s + V_{RR}^t|^2 + 4 \, |S_{LR}^t|^2+ 12
\, |S_{RL}^s|^2}  \, \, .
\label{eq:afbr}
\end{equation}
Finally, the left-right asymmetry reads
\begin{equation}
A_{LR} \, = \, \frac{|V_{LL}^s + V_{LL}^t|^2 - |V_{RR}^s +
V_{RR}^t|^2+ |S_{RL}^t|^2 - |S_{LR}^t|^2 + 3 \,
(|S_{LR}^s|^2-|S_{RL}^s|^2)}{|V_{LL}^s + V_{LL}^t|^2 + |V_{RR}^s +
V_{RR}^t|^2+ |S_{RL}^t|^2 + |S_{LR}^t|^2 + 3 \,
(|S_{LR}^s|^2+|S_{RL}^s|^2)} \, ,
\label{eq:alr}
\end{equation}
which has no dependence on $\Lambda$. Notice that all of these
expressions assume an unpolarized positron beam, and a completely
polarized electron beam, either left- or right-handed. If the
electron beam is not perfectly polarized, but instead has a
percentage of polarization $P_{e^-}$, we can still write
\begin{equation}
\sigma_{P_{e^-}} \, =   \, \sigma_0 \left[ 1-P_{e^-} A_{LR}
\right]
\end{equation}
with $\sigma_0 \, = \, (\sigma_{L}+\sigma_{R})/4$. So if in reality
we only have access at the ILC to beams with +80 \% (- 80 \%)
polarization we could still use them to determine $\sigma_0$ and
$A_{LR}$. If we had access to a positron polarized beam, we could
then write a similar expression for the cross section obtained from
the polarized positrons. Notice that $A_{LR}$ would be different -
the indices left and right would then refer to the positron and not
to the electron.

The most interesting possibility is, of course, when both beams are
polarized, with different percentages, $P_{e^-}$ and $P_{e^+}$. We
could then perform experiments where the four different combinations
of beam polarizations were used. The resulting cross section would
be
\begin{align}
\sigma_{P_{e^-}P_{e^+}} \, = &  \, \frac{1}{4} \biggl[
(1+P_{e^-})(1+P_{e^+}) \sigma_{RR}\, + \, (1-P_{e^-})(1-P_{e^+})
\sigma_{LL}+
\nonumber \\
& (1+P_{e^-})(1 - P_{e^+})  \sigma_{RL} \, + \,
(1-P_{e^-})(1+P_{e^+}) \sigma_{LR} \biggr]\;\;\; .
\end{align}
As such, we would be able to determine $\sigma_{RR}$, $\sigma_{LL}$,
$\sigma_{RL}$ and $\sigma_{LR}$ - and consequently each of the four
four-fermion couplings, $V_{LL}$, $V_{RR}$, $S_{RL}$ and $S_{LR}$.

\section{Results and Discussion}
\label{sec:res}
\begin{figure}[htbp]
  \begin{center}
    \epsfig{file=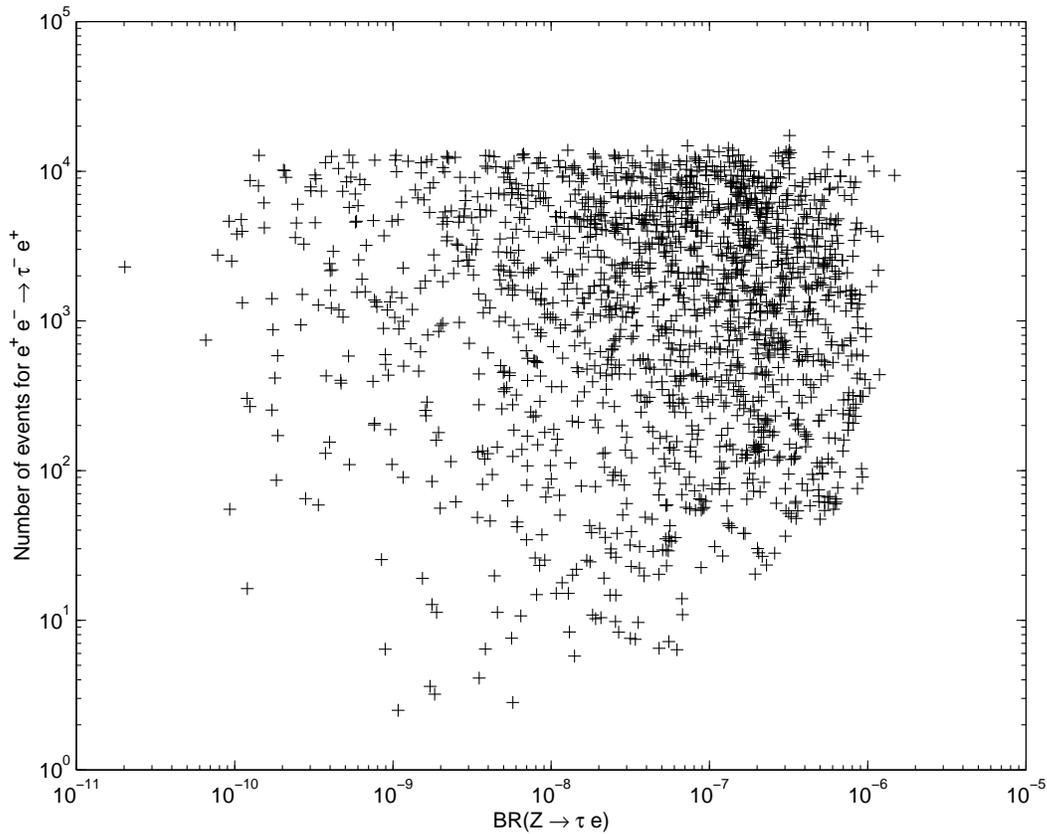,width=14 cm}
    \caption{Number of expected events at the ILC for the reaction $e^+\,e^-\,\rightarrow\,\tau^-\,
    e^+$, with a center-of-mass energy of 1 TeV and a total luminosity of 1 ab$^{-1}$.}
    \label{fig:plot2}
  \end{center}
\end{figure}

\noindent In the previous sections we computed cross sections and
decay widths for several flavour-violating processes. We will now
consider the possibility of their observation at the ILC. To do so
we will use one set of parameters \cite{Moortgat-Pick:2005cw} for
the ILC, i.e., a center-of mass energy of $\sqrt{s} = 1$ TeV and an
integrated luminosity of ${\cal L} = 1$ ab$^{-1}$. At this point we
remark that, other than the experimental constraints on the
flavour-violating decay widths computed in sec.~\ref{sec:dec} (see
table~\ref{tab:dec}), we have no bounds on the values of the
anomalous couplings.
The range of values chosen for each of the coupling constants was
$10^{-4} \leq |a/\Lambda^2| \leq 10^{-1}$, where $a$ stands for a
generic coupling and $\Lambda$ is in TeV. For $a \approx 1$ the
scale of new physics can be as large as 100 TeV. This means that if
the scale for LFV is much larger than 100 TeV, it will not be probed
at the ILC unless the values of coupling constants are unusually
large. The asymmetry plots are not affected by this choice as
explained before.
\begin{table}[h]
\caption{Experimental constraints on flavour-violating decay
branching ratios~\cite{pdg}.}
\begin{ruledtabular}
\begin{tabular}{cc}
 Process & Upper bound
\\
\hline $\tau\,\rightarrow\,e\,e\,e$ & $2.0 \times 10^{-7}$    \\
$\tau\,\rightarrow\,e\,\mu\,\mu$ & $2.0 \times 10^{-7}$    \\
$\tau\,\rightarrow\,\mu\,e\,e$ & $1.1 \times 10^{-7}$   \\
$\tau\,\rightarrow\,\mu\,\mu\,\mu$ & $1.9 \times 10^{-7}$     \\
$\mu\,\rightarrow\,e\,e\,e$ & $1.0 \times 10^{-12}$   \vspace{0.2cm}  \\
  \hline
$Z\,\rightarrow\,e\,\mu$ & $1.7 \times 10^{-6}$   \\
$Z\,\rightarrow\,e\,\tau$ & $9.8 \times 10^{-6}$
 \\
$Z\,\rightarrow\,\tau\,\mu$ & $1.2 \times 10^{-5}$  \\
\end{tabular}
\end{ruledtabular}
\label{tab:dec}
\end{table}
We will therefore generate random values for all anomalous
couplings (four-fermion and $Z$ alike), and discard those
combinations of values of the couplings for which the several
branching ratios we computed earlier are larger than the
corresponding experimental upper bounds from table~\ref{tab:dec}.
This procedure allows for the possibility that one set of
anomalous couplings (the $Z$ or four-fermion ones) might be much
larger than the other. When an acceptable combination of values is
found, it is used in expressions~\eqref{eq:sig4f}-~\eqref{eq:int}
to compute the value of the flavour-violating cross section. In
figure~\ref{fig:plot2} we plot the number of events expected at
the ILC for the process
$e^+\,e^-\,\rightarrow\,\tau^-\,e^+$, in terms of the branching
ratio BR$(Z\,\rightarrow\,\tau\,e)$. To obtain the points shown in
this graph, we demanded that the values of the effective couplings
were such that all of the branching ratios for the decays of the
$\tau$ lepton into three light leptons and
BR$(Z\,\rightarrow\,\tau\,\mu)$ were smaller than the experimental
upper bounds on those quantities shown in table~\ref{tab:dec}. We
observe that, even for fairly small values of the $\tau$
flavour-violating decay branching ratios ($10^{-9}$-$10^{-6}$),
there is the possibility of a large number of events for the
anomalous cross section.

\begin{figure}[htbp]
  \begin{center}
    \epsfig{file=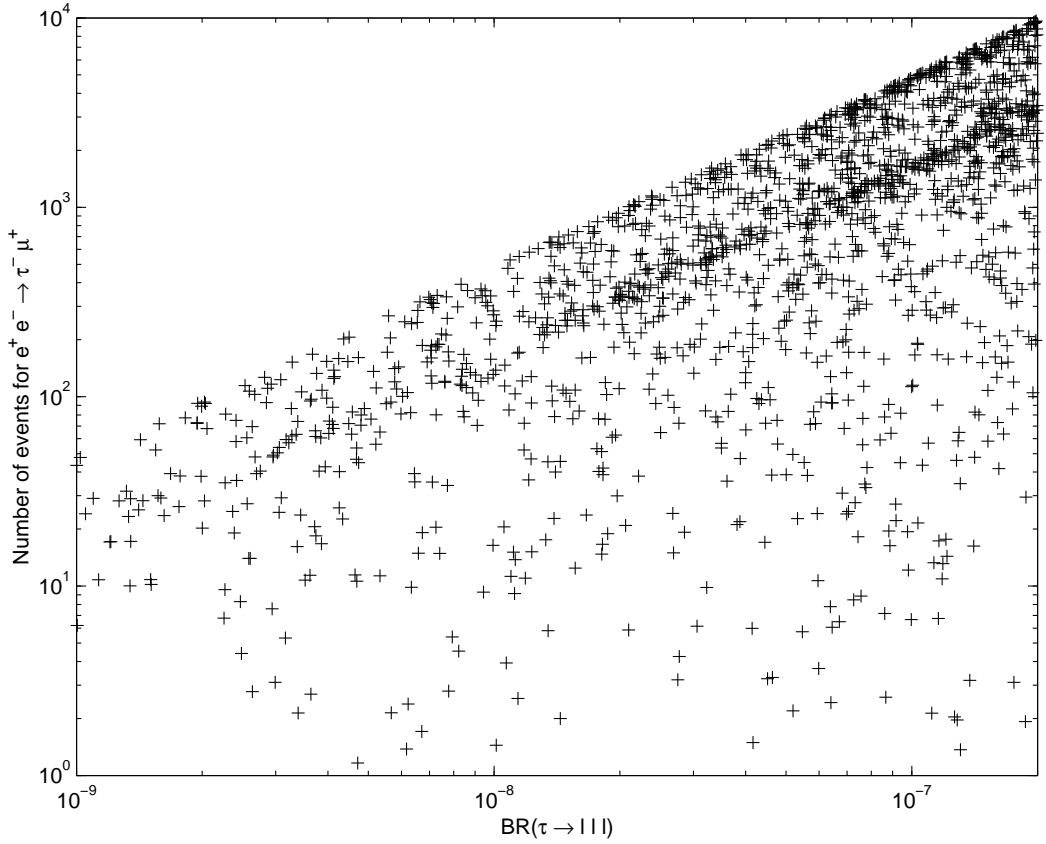,width=14 cm}
    \caption{Number of expected events at the ILC for the reaction $e^+\,e^-\,\rightarrow\,\tau^-\,
    \mu^+$, with a center-of-mass energy of 1 TeV and a total luminosity of 1 ab$^{-1}$.}
    \label{fig:plot3}
  \end{center}
\end{figure}

\begin{figure}[htbp]
  \begin{center}
    \epsfig{file=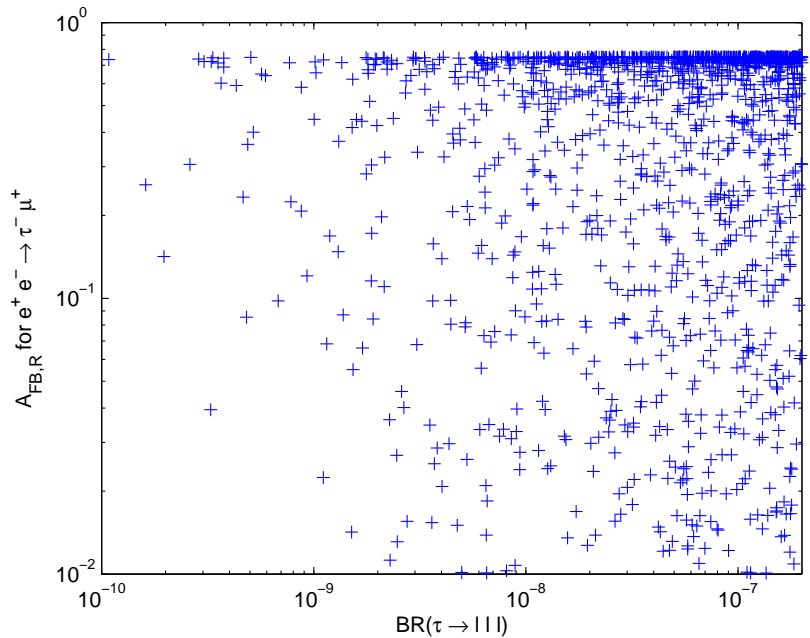,width=12 cm}
    \caption{$A_{FB,R}$ asymmetry for the process $e^+\,e^-\,\rightarrow\,
    \tau^-\,\mu^+$ versus $BR(\tau\,\rightarrow\,l\,l\,l)$}
    \label{fig:asy}
  \end{center}
\end{figure}

By following the opposite procedure - requiring first that the
branching ratios BR$(Z\,\rightarrow\,\tau\,e)$ and
BR$(Z\,\rightarrow\,\tau\,\mu)$ be according to the experimental
values, and letting BR$(\tau\,\rightarrow\,l\,l\,l)$ free, where
$l$ is either an electron or a muon - we obtain the plot shown in
figure~\ref{fig:plot3}. This time we analyse the process
$e^+\,e^-\,\rightarrow\,\tau^-\, \mu^+$, but a similar plot is found
for $e^+\,e^-\,\rightarrow\,\tau^-\, e^+$. The number of events
rises sharply with increasing branching ratio of $\tau$ into three
leptons. It is possible to discern a thin ``band" of events in the
middle of the points of figure~\ref{fig:plot3}, rising linearly with
BR$(\tau\,\rightarrow\,l\,l\,l)$. This ``band" corresponds to events
for which the four-fermion couplings are dominant over $Z^0$ events.
In that case, they dominate both BR$(\tau\,\rightarrow\,l\,l\,l)$
and $\sigma(e^+\,e^-\,\rightarrow\,\tau^-\, \mu^+)$, and the larger
 one is, the larger the other will be - which explains the linear
growth of this subset of points in the plot of
figure~\ref{fig:plot3}. This ``isolated" contribution from the
four-fermion terms is not visible in figure~\ref{fig:plot2} since
the branching ratios of the $Z$ decays are independent of those
same couplings. Finally and for completeness, in
figure~\ref{fig:asy} we show the values of the asymmetry
coefficient $A_{FB,R}$ defined in~\eqref{eq:afbr}, for the process
$e^+\,e^-\,\rightarrow\,\tau^-\, \mu^+$, versus the three-lepton
decay of the $\tau$. A similar plot is obtained for the asymmetry
$A_{FB,L}$. We observe a fairly uniform dependence on the branching
ratio $BR(\tau\,\rightarrow\,l\,l\,l)$, which is to say, on the
values of the four-fermion couplings. However, there is a
significant concentration of ``points" near the maximum allowed
value for this cross section, 0.75.

Finally, we also considered another possible process of LFV, namely
$\gamma\,e^-\,\rightarrow\,\mu^-\,Z^0$. There are three Feynman
diagrams contributing to this process, one of which involving a
quartic vertex which emerges from the effective operators of
eqs.~\eqref{eq:op1} and~\eqref{eq:op2}. This process might occur at
the ILC, if we consider the almost-collinear photons emitted by the
colliding leptons, well described by the so-called equivalent photon
approximation (EPA)~\cite{epa}. An estimate of the cross section for
this process, however, showed it to be much lower than the remaining
ones we considered in this paper. This is due to the EPA introducing
an extra electromagnetic coupling constant into the cross section,
and also to the fact that the final state of this process includes
at least three particles (one of the beam particles ``survives" the
interaction)- thus there is, compared to the other processes which
have only leptons in the final state, an additional phase space
suppression. Notice, however, that an optional upgrade for the ILC
is to have $e\,\gamma$ collisions, with center-of-mass energies and
luminosities similar to those of the $e^+\,e^-$ mode, so this cross
section might become important.

The flavour-violating channels are experimentally interesting, as
they present a final state with an extremely clear signal, which can
be easily identified. The argument is that the final state will
always present two very energetic leptons of different flavour, more
to the point, an electron and a muon. LFV can be seen in one of the
three channels $e^+ e^- \rightarrow \mu^- e^+$, $e^+ e^- \rightarrow
\tau^- e^+$, $e^+ e^- \rightarrow \tau^- \mu^+$ and charge conjugate
channels. The first channel is the best one, with the two leptons
back to back and almost free of backgrounds. For the other
production processes, we may ``select" the decays of the tau that
best suit our purposes: for the second we should take the tau decay
$\tau^- \rightarrow \mu^- \bar{\nu_{\mu}} \nu_{\tau}$, and for the
third process, $\tau^- \rightarrow e^- \bar{\nu_{e}} \nu_{\tau}$.
The branching ratios for both of these tau decays are around 17\%,
so the loss of signal is affordable. The conclusion is that, for
every lepton flavour-violating process, one can always end up with a
final state with an electron and a muon. If the ILC detectors have
superb detection performances for these particles, then the odds of
observing violation of the leptonic number at the ILC, if those
processes do exist, seem reasonable.

Clearly, our prediction that significant numbers of anomalous events
may be produced at the ILC needs to be further investigated
including the effects of a real detector. Notice also that due to
{\em beamstrahlung} effects which reduce the effective beam energy,
the total LFV event rates might be reduced, specially in the case of
the four-fermion cross sections, which increase with $s$. Also, one
must take into account the many different backgrounds that could
mask our signal. And the fact that, even in the best-case-scenario,
only a few thousand events are produced with an integrated
luminosity of 1 ab$^{-1}$, could limit the signal-to-background
ratio. A careful study of the background to LFV processes lies
clearly beyond the scope of this paper. Nevertheless, we will show
that with some very simple cuts most of the background can be
eliminated. Due to the weaker experimental constraints on processes
involving $\tau$ leptons, the most promising LFV reactions at the
linear collider are $\mu e$ and $\mu \tau$ production. For
illustrative purposes we will study the backgrounds to the LFV
process $e^+ e^- \rightarrow \tau^+ \, e^- \rightarrow \mu^+ \, e^-
\, \nu_{\mu} \, \bar{\nu}_{\tau}$. The main sources of background to
this process are $e^+ e^- \rightarrow \mu^+ \, e^- \, \nu_{\mu} \,
\bar{\nu}_{e}$ and $e^+ e^- \rightarrow \tau^+ \, \tau^- \rightarrow
\mu^+ \, e^- \, \nu_{\mu} \, \bar{\nu}_{e} \, \nu_{\tau} \,
\bar{\nu}_{\tau}$. The cross section to the background process $e^+
e^- \rightarrow \mu^+ \, e^- \, \nu_{\mu} \, \bar{\nu}_{e}$ was
calculated using WPHACT~\cite{WPHACT} and confirmed using
RacoonWW~\cite{RacoonWW}. The cross section for the remaining
background was evaluated using PYTHIA~\cite{PYTHIA}. In $e^+ e^-
\rightarrow \tau^+ \, e^- \rightarrow \mu^+ \, e^- \, \nu_{\mu} \,
\bar{\nu}_{\tau}$ the electron is produced in a two body final
state. Therefore its energy is approximately half of the center of
mass energy. Furthermore, if $\theta_e$ is the angle between the
electron and the beam, then the transverse momentum of the electron
is $p_T=\sqrt{s}/2 \, \sin (\theta_e)$. This means that a cut in
$\theta_e$ implies a cut in $p_T$. The main contribution to this
cross section comes from the four-fermion interaction. There are no
propagators involved and consequently the dependence in $\theta_e$
(and in $p_T$) is very mild. This can be seen from the
expression~\eqref{eq:4ftot} in the appendix. Making all coupling
constants $V_{ij}$ and $S_{ij}$ equal, it can be shown that a 10
degree cut will reduce the cross section by 2 \% while a 60 degree
cut will reduce it only by 58 \%.
\begin{table}[th]
\begin{center}
\begin{tabular}{ccccccc}\hline\hline \\
 Cut in $\theta_e$ (degrees) & 10 & 20 & 30 & 40 & 50 & 60 \\ & & & & & \\ \hline \\
 $e^+ e^- \rightarrow
\tau^+ \, e^- \rightarrow  \mu^+ \, e^- \, \nu_{\mu} \,
\bar{\nu}_{\tau}$ \qquad \quad
 & 4.9 & 4.6 & 4.1 & 3.5 & 2.8 & 2.1 \vspace{0.2cm} \\
$e^+ e^- \rightarrow \mu^+ \, e^- \, \nu_{\mu} \, \bar{\nu}_{e}$
\qquad \quad
 & 68.2 & 26.3 & 10.8 & 4.4 & 1.6 & 0.5 \vspace{0.2cm} \\
$e^+ e^- \rightarrow \tau^+ \, \tau^- \rightarrow \mu^+ \, e^- \,
\nu_{\mu} \, \bar{\nu}_{e} \, \nu_{\tau} \, \bar{\nu}_{\tau}$ \qquad
\quad
 & 1.3 & 0.8 & 0.3 & 0.2 & 0.06 & 0.01 \vspace{0.2cm}
\\\hline\hline\hline
\label{tab:ggq}
\end{tabular}
\caption{Cross sections (in fbarn) for the LFV signal and most
relevant backgrounds to that process for several values of the angle
cut between the the outgoing electron and the beam axis.}
\end{center}
\end{table}
In table I we show the cross sections for the signal and for the
backgrounds as a function of a cut in $\theta_e$ and a
corresponding cut in $p_T$. For the signal we start with a cross
section of 5 fbarn when no cuts are applied. Due to the mild
dependence on $\theta_e$, a cut of 60 degrees will make the signal
well above background. A further cut on the energy of the electron
could be applied, say $E_e > 300 \,  GeV$. This would not affect
the signal but will reduce the background even further. All
calculations were performed at tree level with Initial State
Radiation and Final State Radiation turned off. Another
possibility for background reduction would be to use the
polarisation of the beams, a method known to be very efficient.
Notice, however, that this procedure might affect the extraction
of four-fermion couplings from polarised beam experiments - if the
signal is observed only for certain combinations of beam
polarisations, it could happen that only certain couplings, or
combinations thereof, can be measured.

Finally, some comments on the dependence of these results vis-a-vis
expected improvements on the measurements of the LFV branching
ratios of table~\ref{tab:dec}. Could it be that future experiments
would tighten the constraints so much that there was no room
available for discovery? Tau physics at BABAR and BELLE has provided
the best limits so far on LFV involving the $\tau$ lepton. The
combined results from BABAR and BELLE on $\tau \rightarrow l \gamma$
are now reaching the level of $10^{-8}$ and will be close to just a
few $10^{-8}$ by 2008~\cite{BB}. More important to us are the decays
$\tau \rightarrow lll$, due to the constraints imposed on the four
fermion operators. The latest results on $Br(\tau \rightarrow lll)$
from BABAR and BELLE are of the order of $10^{-7}$, with less than
$100 \, fb^{-1}$ of data analysed. A value of the order of a few
$10^{-8}$ is expected when all data is taken into account~\cite{FS}.
Other planned experiments like MEG or SINDRUM2 (see ~\cite{Nicolo})
will provide much more precise results for both $\mu \rightarrow e
\gamma$ and $\mu \, e$ conversion, respectively. However, those
results will not constrain any further the four-fermion couplings.
The current limit $Br(\mu \rightarrow e\,e\,e ) < 10^{-12}$ at 90\%
CL~\cite{Bellgardt} already excludes the possibility of finding LFV
in the $\mu\, e\,e\,e$ coupling. This limit will be improved by the
Sundrum experiment (see~\cite{Nicolo}).
Another possibility is the GigaZ option for the ILC, which
probably would be earlier than an energy upgrade to 1 $TeV$.
Again, the limits on the LFV branching ratios of the Z boson would
be improved~\cite{Wilson} but the bounds on the four-fermion
couplings would not be affected.
Lastly, LFV searches will also take place at the LHC. Preparatory
studies on the LFV decay $\tau \rightarrow \mu\, \mu\, \mu$ are
being conducted by CMS~\cite{CMSLFV}, ATLAS~\cite{LHCLFV} and also
by LHCb~\cite{LHCbLFV}. During the initial low luminosity runs
(10-30 $fb^{-1}$/year) for 2008-2009, searches for this decay may
be possible. So far the limits predicted are only slightly better
than the known limits from the B-factories. Therefore, in the
foreseeable future, the constraints on the four fermion $\tau$
couplings arising from the branching ratios of table~\ref{tab:dec}
could go down one order of magnitude, to be of the order of
$10^{-8}$. Accordingly, and repeating the calculations that led to
figs.~\ref{fig:plot2} and~\ref{fig:plot3}, the maximum number of
events expected at the ILC also goes down by one order of
magnitude, to about 1000 events. Given the discussion on
backgrounds above, we expect that detection of LFV at the ILC
would still be possible, although harder.

\vspace{0.25cm} {\bf Acknowledgments:} Our thanks to Jose Wudka for
many useful discussions. We benefitted from many conversations with
Pedro Teixeira Dias and Ricardo Gon\c{c}alo on several experimental
subjects. A very special thanks to Nuno Castro for checking with
WPHACT the values we obtained with RacoonWW, and, with Filipe
Veloso, for having given us the PYTHIA code for tau production. Our
further thanks to Augusto Barroso, Ant\'onio Onofre, PTD and RG for
a careful reading of the manuscript. This work is supported by
Funda\c{c}\~ao para a Ci\^encia e Tecnologia under contract
POCI/FIS/59741/2004. P.M.F. is supported by FCT under contract
SFRH/BPD/5575/2001. R.S. is supported by FCT under contract
SFRH/BPD/23427/2005. R.G.J. is supported by FCT under contract
SFRH/BD/19781/2004.

\appendix

\section{Single top production via gamma-gamma collisions}

In section~\ref{sec:gamma} we argued that the couplings
corresponding to the operators of eqs.~\eqref{eq:op4} were extremely
limited in size by the existing experimental data for the branching
ratios of the decays $l_h\,\rightarrow\,\l_l\,\gamma$. In fact, we
even showed that the cross sections for the processes
$\gamma\,\gamma\,\rightarrow\,l_h\,l_l$, eq.~\eqref{eq:ggl}, were
directly proportional to those branching ratios, and their values at
the ILC were predicted to be exceedingly small. It is easy to
understand, though, that we can define operators analogous to those
of eqs.~\eqref{eq:op4} for quarks instead of leptons. In particular,
we can consider flavour-changing operators involving the top quark,
which would describe decays such as $t\,\rightarrow\,u\,\gamma$ or
$t\,\rightarrow\,c\,\gamma$ - and these decay widths have not yet
been measured. More importantly, their values may vary immensely,
depending on the model one uses to calculate them. According
to~\cite{juan}, the branching ratios for these decays range from
their SM value of $\sim\,10^{-16}$ (for the $u$ quark),
$\sim\,10^{-12}$ (for the $c$ quark) to $\sim\,10^{-6}$ (for both
quarks) in supersymmetric models with R-parity violation. The total
top quark width being also a lot larger than the tau's or the
muon's, it seems possible that the cross section for single top
production via flavour-violating photon-photon interactions presents
us with observable values.

The corresponding calculation is altogether identical to the one we
presented for the leptonic case. We find an expression for the width
of the anomalous decay $t\,\rightarrow\,q\,\gamma$ similar to that
of eq.~\eqref{eq:wid},
\begin{eqnarray}
\Gamma(t\,\rightarrow \,q\,\gamma)&=&\frac{m_t^3}{64 \pi
\Lambda^4}\biggl[
 m_t^2(|\alpha^{\gamma R}_{qt}+\alpha^{{\gamma R}*}_{tq}|^2+|\alpha^{\gamma L}_{qt}+
 \alpha^{{\gamma L}*}_{tq}|^2) + 16v^2\,(|\beta^{\gamma}_{qt}|^2+|\beta^{\gamma}_{tq}|^2)\nonumber\\
&+&  8\,m_t\,v\: \mbox{Im}(\alpha^{\gamma
R}_{tq}\beta^{\gamma}_{tq}-\alpha^{\gamma R}_{qt}\beta^{\gamma
*}_{tq}-\alpha^{\gamma L}_{qt}\beta^{\gamma *}_{qt}+\alpha^{\gamma L}_{tq}\beta^{\gamma}_{tq})
\biggr] \, \, ,
\end{eqnarray}
with {\em new} couplings $\alpha$ and $\beta$ (we re-emphasize that
these new couplings are not in the least constrained by the
arguments we used in section~\ref{sec:gamma}) and
$q\,=\,\{u\,,\,c\}$. Likewise, considering that the top quark's
charge is $2/3$ and the quarks have three colour degrees of freedom,
we may rewrite the analog of eq.~\eqref{eq:ggl} as
\begin{equation}
\frac{d\sigma(\gamma\,\gamma\,\rightarrow \,t\,\bar{q})}{d t} \;
\;=\;\; -\,\frac{16\pi\alpha\,F_{\gamma\gamma}}{3\,{m_t}^3\,s\,
    {({m_t}^2 - t ) }^2\,t\,{( {m_t}^2 - u ) }^2\,u}\;\Gamma(t\,\rightarrow
    \,q\,\gamma)\;\;\;,
    \label{eq:ggtq}
\end{equation}
with $F_{\gamma\gamma}$ given by an expression identical to
eq.~\eqref{eq:fgg}, with the substitution
$m_h\,\leftrightarrow\,m_t$. With a top total width of about 1.42
GeV and for $\sqrt{s}$ equal to 1 TeV, this expression can be
integrated in $t$ (with a $p_T$ cut of 10 GeV on the final state
particles, to prevent any collinear singularities) and the total
cross section estimated to be of the order
\begin{align}
\sigma(\gamma\,\gamma\,\rightarrow \,t\,\bar{q})\;\sim&
\;\;\;90\,\times\,\mbox{BR}(t\,\rightarrow \,q\,\gamma)
\;\;\mbox{pb} \;\;\; .
\end{align}
We see a considerable difference vis-a-vis the predicted leptonic
cross sections, from eqs.~\eqref{eq:crl} - this one is much larger.
To pass from the photon-photon cross section to an electron-positron
process, we apply the standard procedure: use the equivalent photon
approximation~\cite{epa} to provide us with the probability of an
electron/positron with energy $E$ radiating photons with a fraction
$x$ of E and integrate eq.~\eqref{eq:ggtq} over $x$. For recent
studies of photon-photon collisions at the ILC, see for
instance~\cite{epailc}. The numerical result we found for the single
top production cross section is
\begin{equation}
\sigma(e^+\,e^-\,\rightarrow \,e^+\,e^-\,t\,\bar{q})\;\;=
\;\;1.08\,\times\,\mbox{BR}(t\,\rightarrow \,q\,\gamma)
\;\;\mbox{pb} \;\;\; . \label{eq:ggtop}
\end{equation}
For an integrated luminosity of about 1 ab$^{-1}$, this gives us
about one event observed at the ILC for branching ratios of
$t\,\rightarrow \,q\,\gamma$ near the maximum of its theoretical
predictions~\cite{note3}, $\sim\,10^{-6}$. Clearly, this result
means that this process should not be observed at the ILC, even in
the best case scenario. However, in the event of non-observation,
eq.~\eqref{eq:ggtop} could be useful to impose an indirect limit on
the branching ratio $\mbox{BR}(t\,\rightarrow \,q\,\gamma)$. Several
authors have studied single top production in $e^+\,e^-$
collisions~\cite{Bar-Shalom:1999iy,tcilc}. For gamma-gamma
reactions, single top production at the ILC in the framework of the
effective operator formalism may has been studied in~\cite{ggtop},
and for specific models, such as SUSY and technicolor, in
ref.~\cite{epailc}.

\section{Total cross section expressions}
\label{sec:cross2} \noindent We write the amplitude for the four
fermion cross sections in two parts. One for the $s$ channel and
the other one for the $t$ channel. In doing so we are generalizing
the four fermion lagrangian which for a gauge theory has equal
couplings for both $s$ and $t$ channels. For the $s$ channel the
amplitude reads
\begin{equation}
T_{ij}^s \, = \, \frac{1}{\Lambda^4} \left[ V_{ij}^s \, (\bar{v}_e
\gamma_{\alpha} \gamma_i u_e) (\bar{u}_{l_h} \gamma^{\alpha}
\gamma_j v_{l_l}) \, + \, \, S_{ij}^s \, (\bar{v}_e \gamma_i u_e)
(\bar{u}_{l_h} \gamma_j v_{l_l}) \right]
\end{equation}
while for the t channel we have
\begin{equation}
T_{ij}^t \, = \,  -\,\frac{1}{\Lambda^4} \left[ V_{ij}^t \,
(\bar{u}_{l_h} \gamma_{\alpha} \gamma_i u_e) (\bar{v}_e
\gamma^{\alpha} \gamma_j v_e) \, + \, \, S_{ij}^t \, (\bar{u}_{l_h}
\gamma_i u_e) (\bar{v}_e \gamma_j v_e) \right]
\end{equation}
with $i,j = L,R$. With these definitions we can write
\begin{align}
|T (e_L^- e_L^+ \rightarrow {l_h}_L^- {l_l}_L^+)|^2 \, = &
\,\frac{1}{\Lambda^4} (4 u^2 |V_{LL}^s + V_{LL}^t|^2)
\nonumber \\
|T (e_R^- e_R^+ \rightarrow {l_h}_R^- {l_l}_R^+)|^2 \, = & \,
\frac{1}{\Lambda^4} (4 u^2
|V_{RR}^s + V_{RR}^t|^2) \nonumber \\
|T (e_L^- e_L^+ \rightarrow {l_h}_R^- {l_l}_R^+)|^2 \, = & \,
\frac{1}{\Lambda^4}\,t^2 \,|S_{RL}^t|^2
\nonumber \\
|T (e_R^- e_R^+ \rightarrow {l_h}_L^- {l_l}_L^+)|^2 \, = & \,
\frac{1}{\Lambda^4} t^2 \,|S_{LR}^t|^2
\nonumber \\
|T (e_L^- e_R^+ \rightarrow {l_h}_L^- {l_l}_R^+)|^2 \, = & \,
\frac{1}{\Lambda^4} \, s^2 \,|S_{LR}^s|^2
\nonumber \\
|T (e_R^- e_L^+ \rightarrow {l_h}_R^- {l_l}_L^+)|^2 \, = & \,
\frac{1}{\Lambda^4} \, s^2 \, |S_{RL}^s|^2
\end{align}
and to obtain the expressions when only the $t$ or $s$ channels are
present, you just have to set the $s$ couplings or the $t$
couplings, respectively, equal to zero. $u$, $t$ and $s$ are the
Mandelstam variables defined in the usual way.

The cross sections for polarized electron and positron beams with
no detection of the polarization of the final state particles were
given in eq.~\eqref{eq:sig4f}. The International Linear Collider
will have a definite degree of polarization that will depend on
the final design of the machine. For longitudinally polarized
beams the cross section can be written as
\begin{align}
\frac{d \sigma_{P_{e^-}P_{e^+}}}{dt} \, = &  \, \frac{1}{4}
\biggl[ (1+P_{e^-})(1+P_{e^+}) \frac{d \sigma_{RR}}{dt} \, + \,
(1-P_{e^-})(1-P_{e^+}) \frac{d \sigma_{LL}}{dt}+
\nonumber \\
& (1+P_{e^-})(1 - P_{e^+}) \frac{d \sigma_{RL}}{dt} \, + \,
(1-P_{e^-})(1+P_{e^+}) \frac{d \sigma_{LR}}{dt} \biggr]
\end{align}
where $\sigma_{RL}$ corresponds to a cross section where the
electron beam is completely right-handed polarized ($P_{e^-}=+1$)
and the positron beam is completely left-handed polarized
($P_{e^+}=-1$). This reduces to the usual averaging over spins in
the case of totally unpolarized beams. For the general expression
for polarized beams, as well as a study on all the advantages of
using those beams, see~\cite{Moortgat-Pick:2005cw}.

In the main text we presented expressions for the differential cross
sections. For completeness we now present the formulae for the total
cross sections. For the four-fermion case, the expressions have a
very simple dependence on the $p_T$ cut one might wish to apply, so
we exhibit it. The quantity $x=\sqrt{1-4p_T^2/s}$, with $p_T$ being
the value of the minimum transverse momentum for the heaviest
lepton, gives us an immediate way of obtaining these cross sections
with a cut on the $p_T$ of the final particles. The total cross
section is obviously the sum over all polarized ones, which gives us
\begin{equation}
\sigma \, =  \, \frac{s \, x \, (3+x^2)}{768 \pi \Lambda^4} \left( 4
\, |V_{LL}^s + V_{LL}^t|^2 + |S_{RL}^t|^2  + 4\,
 |V_{RR}^s + V_{RR}^t|^2 + |S_{LR}^t|^2 \right)
+ \, \frac{s \, x}{64 \pi \Lambda^4} \left( |S_{LR}^s|^2 +
|S_{RL}^s|^2 \right) \, . \label{eq:4ftot}
\end{equation}
As explained in the main text, the cross sections for processes
$(1,2)$ are obtained from eq.~\eqref{eq:4ftot} by setting all of the
``s" couplings equal to the ``t" ones, and, for process $(3)$, by
setting the ``t" couplings to zero.

For the remaining cross section expressions we imposed no $p_T$ cut
on any of the final particles. The total cross section for the $Z$
couplings is given by, for processes $(1,2)$,
\begin{align}
\sigma_Z^{(1,2)} (e^- e^+ \rightarrow l_h l_l) \;=&\;\; \frac{
v^2}{192 \,\pi\,\Lambda^4\,M_z^2\, s^2\,(M_z^2-s)^2 \, (M_z^2+s) }
\left[ F_3 (g_A) |\eta_{lh}|^2 \right.
\nonumber \vspace{0.3cm} \\
& \; \left. + F_3 (-g_A) |\eta_{hl}|^2 + F_4 (g_A) |\theta_{L}|^2 +
F_4 (-g_A) |\theta_{R}|^2  \right] \, \, ,
\end{align}
with
\begin{align}
F_3 (g_A) \;=&\;\; 6 \,s\,M_z^2\,(M_z^4-s^2) \log \left(
\frac{M_z^2+s}{M_z^2}\right) \left[ (g_A-g_V)^2 M_z^4 -2 (g_A^2+g_A
g_V+g_V^2) s M_z^2 - (g_A^2+g_V^2) s^2 \right]
\nonumber \vspace{0.3cm} \\
& \; - s^2\,M_z^2\, (M_z^2+s) \left[ 6(g_A-g_V)^2 M_z^4 -3 (7g_A^2-2
g_A g_V+7 g_V^2) s M_z^2 + 2 (7g_A^2+3 g_A g_V+7 g_V^2) s^2 \right]
\nonumber \\
& \\
F_4 (g_A) \;=&\;\; 48 v^2 (M_z^4-s^2) (M_z^2+s) M_z^4 \log \left(
\frac{M_z^2}{M_z^2+s}\right) (g_A-g_V)^2 +8 v^2 s \left[ 3
(g_A-g_V)^2 M_z^6 (2M_z^2 + s) \right.
\nonumber \vspace{0.3cm} \\
& \left. - (5 g_A^2 - 18 g_A g_V+ 5 g_V^2) s^2 M_z^4 - 5
(g_A^2+g_V^2) s^3 M_z^2 + 3 (g_A^2+g_V^2) s^4 \right]
 \nonumber \; ,
\end{align}
with interference terms
\begin{align}
\sigma_{int}^{(1,2)} \;=&\;\; -\,\frac{ v^2}{48 \, \Lambda^4\,\pi
\,(s-M_z^2)\,s^2 } \biggr\{
\nonumber \vspace{0.3cm} \\
& s \biggr[ \left( g_A -g_V \right) \left\{
(12M_z^4+6sM_z^2-14s^2)\mbox{Re} \left( \theta_L V_{LL}^*
\right)-s^2\mbox{Re} \left( \theta_R S_{RL}^* \right) \right\}
\nonumber \vspace{0.3cm} \\
& \, \,  -\left( g_A +g_V \right) \left\{
(12M_z^4+6sM_z^2-14s^2)\mbox{Re} \left( \theta_R V_{RR}^*
\right)-s^2\mbox{Re} \left( \theta_L S_{LR}^* \right) \right\}
\biggl]
\nonumber \vspace{0.3cm} \\
& +\,3\,(M_z^2-s) \biggr[ \left( g_A -g_V \right) \left\{
(4M_z^4+8sM_z^2-4s^2)\mbox{Re} \left( \theta_L V_{LL}^*
\right)-s^2\mbox{Re} \left( \theta_R S_{RL}^* \right) \right\}
\nonumber \vspace{0.3cm} \\
& \, \,  -\left( g_A +g_V \right) \left\{
(4M_z^4+8sM_z^2-4s^2)\mbox{Re} \left( \theta_R V_{RR}^*
\right)-s^2\mbox{Re} \left( \theta_L S_{LR}^* \right) \right\}
\biggl]
\nonumber \vspace{0.3cm} \\
&\log \left( \frac{M_z^2}{M_z^2+s}   \right) \biggl\} \;\;\; .
\end{align}
Finally, for process $(3)$, we have
\begin{equation}
\sigma_Z^{(3)}  \;=\; \frac{\left( g_V^2 +g_A^2 \right) v^2
s}{192\,\pi\, \Lambda^4\,(M_z^2-s)^2 }  \left[ 8 \left(
|\theta_L|^2+|\theta_R|^2 \right) v^2 + \left(
|\eta_{hl}|^2+|\eta_{lh}|^2 \right) s \right] \, \, ,
\end{equation}
and
\begin{align}
\sigma_{int}^{(3)} \;= &\; \frac{s v^2}{24 \pi (s-M_z^2) \,
\Lambda^4} \left[  \left( g_V +g_A \right) \mbox{Re} \left(\theta_R
V_{RR}^* \right) + \left( g_V -g_A \right) \, \mbox{Re} \left(
\theta_L V_{LL}^*\right) \right] \;\;\; .
\end{align}

\section{Numerical values for decay widths and cross sections}

We present here numerical values for the several decay widths and
cross sections given in the text.  We have set, in the following
expressions, $\Lambda$ equal to 1 TeV, the dependence in $\Lambda$
being trivially recovered if we wish a different value for it.

\begin{align}
\mbox{BR}_{4f} (\mu \rightarrow l l l) &=\;  2.3 \times 10^{-4}
(|S_{LR}|^2+|S_{RL}|^2+4(|V_{LL}|^2+|V_{RR}|^2)) \vspace{0.4cm}  \nonumber \\
\mbox{BR}_{4f} (\tau \rightarrow l l l) &=\; 4.0 \times 10^{-5}
(|S_{LR}|^2+|S_{RL}|^2+4(|V_{LL}|^2+|V_{RR}|^2))\vspace{0.4cm}\nonumber \\
\mbox{BR}_Z (\mu \rightarrow l l l) &=\; 8.2 \times 10^{-4} \left(
|\theta_L|^2+|\theta_R|^2 \right)  + 2.5 \times 10^{-7} \mbox{Re}
\left(
\eta_{lh} \theta_L^* +  \eta_{hl} \theta_R \right)\vspace{0.4cm}\nonumber \\
\mbox{BR}_Z (\tau \rightarrow l l l) &=\; 1.4 \times 10^{-4} \left(
|\theta_L|^2+|\theta_R|^2 \right)  + 7.3 \times 10^{-7} \mbox{Re}
\left( \eta_{lh} \theta_L^* +  \eta_{hl} \theta_R
\right)\vspace{0.4cm}\nonumber \\
\mbox{BR}_{int} (\mu \rightarrow l l l) &=\; -1.4 \times 10^{-3}
\mbox{Re} \left( \theta_L V_{LL}^*\right) + 1.1 \times 10^{-3}
\mbox{Re} \left( \theta_R V_{RR}^* \right) \vspace{0.4cm}
\nonumber  \\
& \;   +\,1.7 \times 10^{-7} Re \left( \eta_{lh} V_{RR}^*  \right)
-2.1 \times 10^{-7} \mbox{Re} \left( \eta_{hl}
V_{LL} \right)\vspace{0.4cm}\nonumber \\
\mbox{BR}_{int} (\tau \rightarrow l l l) &=\; -2.4 \times 10^{-4} Re
\left( \theta_L V_{LL}^* \right) + 1.9 \times 10^{-4} \mbox{Re}
\left( \theta_R V_{RR}^* \right) \vspace{0.4cm}
\nonumber  \\
 & \; +\, 4.8 \times 10^{-7} Re \left( \eta_{lh}
V_{RR}^*  \right)  -6.0 \times 10^{-7} \mbox{Re} \left( \eta_{hl}
V_{LL} \right)\;\;\; .
\end{align}

\begin{align}
\mbox{BR} (Z \rightarrow l l)     &=\;   2.3 \times 10^{-5}
(|\eta_{hl}|^2+|\eta_{lh}|^2) + 6.7 \times 10^{-4}
(|\theta_{L}|^2+|\theta_{R}|^2) \nonumber \\
\mbox{BR} (Z \rightarrow \mu l)     &=\;  Br (Z \rightarrow l l)-
2.0 \times 10^{-7} \mbox{Re} \left( \theta_L \eta_{hl}+ \theta_R
\eta_{lh}^* \right) \nonumber \\
\mbox{BR} (Z \rightarrow \tau l)     &=\;   Br (Z \rightarrow l l) -
2.4 \times 10^{-6} \mbox{Re} \left( \theta_L \eta_{hl}+ \theta_R
\eta_{lh}^* \right) \;\;\;.
\end{align}

For the cross sections, taking $\sqrt{s}= 1$ TeV and imposing a
cut of 10 GeV on the $p_T$ of the particles in the final state, we
have (in picobarn):
\begin{align}
\sigma_{4f}^{(1,2)} (e^- e^+ \rightarrow l l)  &=\; 2.58 \left(
|S_{LR}|^2+|S_{RL}|^2 \right) + 10.33 \left( |V_{LL}|^2+|V_{RR}|^2
 \right)\nonumber \\
\sigma_{4f}^{(3)} (e^- e^+ \rightarrow l l)  &=\; 1.94 \left(
|S_{LR}|^2+|S_{RL}|^2 \right) + 2.58 \left(
|V_{LL}|^2+|V_{RR}|^2 \right)\nonumber \\
\sigma_Z^{(1,2)} (e^- e^+ \rightarrow l l) &=\; 1.0 \times 10^{-2}
|\eta_{lh}|^2 +9.7 \times 10^{-3} |\eta_{hl}|^2 + 5.7 \times 10^{-2}
\left( \theta_{L}|^2 +  |\theta_{R}|^2  \right)\nonumber \\
\sigma_Z^{(3)} (e^- e^+ \rightarrow l l)  &=\; 1.6 \times 10^{-4}
\left( |\theta_L|^2+|\theta_R|^2 \right) + 6.7 \times 10^{-4} \left(
|\eta_{hl}|^2+|\eta_{lh}|^2 \right)\nonumber \\
\sigma_{int}^{(1,2)} (e^- e^+ \rightarrow l l) &=\; 0.70 \,
\mbox{Re} \left( \theta_L V_{LL}^* \right) + 0.19 \, \mbox{Re}
\left( \theta_L S_{RL}^* \right) - 0.56 \, \mbox{Re} \left( \theta_R
V_{RR}^* \right) - 0.24 \, \mbox{Re}
\left( \theta_R S_{LR}^* \right)\nonumber \\
\sigma_{int}^{(3)} (e^- e^+ \rightarrow l l)  &= \; - 2.6 \times
10^{-2} \, \mbox{Re} \left( \theta_R V_{RR}^* \right) +3.2 \times
10^{-2} \,  \mbox{Re} \left( \theta_L V_{LL}^* \right) \;\;\; .
\end{align}


\begin{thebibliography}{99}
\bibitem{note1} The tensor operators were eliminated using Fierz
transformations. A tensor exchange can thus be hidden in the vector
and scalar operators.
\bibitem{note2} This cancelation occurs even if we consider the
case where all the leptons have masses.
\end{thebibliography}
\end{document}